\documentclass[a4paper,usenatbib]{mn2e}

\usepackage{hyperref}
\usepackage{color}
\usepackage{amsfonts}
\usepackage{amssymb}
\usepackage{amsmath}
\usepackage{graphicx}
\usepackage[totalwidth=506pt,totalheight=680pt]{geometry}

\newcommand{\adsurl}[1]{\href{#1}{ADS}}
\providecommand{\url}[1]{\href{#1}{#1}}

\newcommand{\avg}[1]{\ensuremath{\langle \,#1\, \rangle}}

\newcommand{\eqn}[1]{equation~\eqref{#1}}

\newcommand{\dd}{\mathrm{d}}
\newcommand{\be}{\begin{equation}}
\newcommand{\ee}{\end{equation}}

\title[Void profiles]
      {Density and velocity profiles around cosmic voids}

\author[E. Massara and R.~K.~Sheth]{%
Elena Massara$^{1,2,3}$\thanks{E-mail: emassara@flatironinstitute.org} 
\& Ravi K.~Sheth$^{4,5}$\thanks{E-mail: shethrk@physics.upenn.edu}\\ 
 $^1$ Center for Computational Astrophysics, Flatiron Institute, 
162 5th Avenue, 10010, New York, NY, USA \\
 $^2$ Berkeley Center for Cosmological Physics, 
 University of California, Berkeley, CA 94720\\
$^3$ SISSA - International School For Advanced Studies,
      Via Bonomea, 265 34136 Trieste, Italy\\
 $^4$ Center for Particle Cosmology, University of Pennsylvania, 
      209 S. 33rd St., Philadelphia, PA 19104, USA\\
 $^5$ The Abdus Salam International Center for Theoretical Physics,
      Strada Costiera, 11, Trieste 34151, Italy}

\begin{document}

\pagerange{\pageref{firstpage}--\pageref{lastpage}}

\maketitle 

\label{firstpage}

%

\begin{abstract}
  We study the evolution of the cross-correlation between voids and the mass density field -- i.e. of void profiles.  We show that approaches based on the spherical model alone miss an important contribution to the evolution on large scales of most interest to cosmology:  they fail to capture the well-known fact that the large-scale bias factor of conserved tracers evolves.  We also show that the operations of evolution and averaging do not commute, but this difference is only significant within about two effective radii.  We show how to include a term which accounts for the evolution of bias, which is directly related to the fact that voids move.  The void motions are approximately independent of void size, so they are more significant for smaller voids that are typically more numerous.  This term also contributes to void-matter pairwise velocities: including it is necessary for modeling the typical outflow speeds around voids.   It is, therefore, important for void redshift space distortions.  Finally, we show that the excursion set peaks/troughs approach provides a useful, but not perfect framework for describing void profiles and their evolution.  
\end{abstract}



\section{Introduction}

The abundance and spatial distribution of voids depends on the nature of the initial conditions, the expansion history of the universe, and the nature of gravity \citep{svdw04,ms09}.  As is true for halos, predictive models of this dependence have three parts.  The first is a description of the gravitational physics of void formation \citep{ddglp93,smt01}; the second incorporates this into a statistical treatment aimed at predicting halo or void abundances, typically from knowledge of the initial fluctuation field \citep{svdw04,pls12,psd13,scs13}; and the third extends this treatment to also describe the spatial distribution of these objects and its evolution \citep{dcss10}.  The initial and evolved fields are often refered to as Lagrangian and Eulerian, respectively.  

The statistical model for the abundances typically identifies those protohalo or protovoid patches in the initial conditions which are destined to become halos and voids.  Since the abundances are predicted from the initial Lagrangian field, whereas the spatial distribution of the fully formed halos and voids is measured in the final evolved Eulerian field, the description of the spatial distribution is done in two steps:  the first describes how the spatial distribution of the initial patches is biased with respect to the initial fluctuation field, and the second describes how this bias is modified when the evolved halo distribution is compared to the evolved matter field.  

In this paper, we first ask if we can model the evolution of void density profiles without any prejudice about how voids form.  We show that naive application of the simplest spherical evolution model does not describe the evolution of void-matter correlation, because voids move.  We then build a more elaborate model which is more accurate.  Finally, we check if the excursion set peaks approach is able to provide a useful framework for describing how matter is initially arranged around protohalo or protovoid centers, showing that it is reasonably accurate.  When coupled with our evolution model, this provides a complete framework for describing the void-matter cross correlation function.

Such a framework is of interest for many cosmological purposes:  Void density profiles are promising for constraining neutrino masses \citep{Massara:2015}, and modeling void velocity profiles allows one to perform redshift space distortion analyses with voids \citep{Hamaus:2015,Cai:2016,Achitouv:2016,Hawken:2016,Nadathur:2018}.  In addition, such a framework is important for interpreting weak-lensing measurements around voids, which are expected to play an important role in the test of theories of gravity \citep{Cai:2014,Baker:2018}.  These studies are motivated in part by the fact that underdensities or voids in the galaxy distribution or lensing signal are now being measured in significant numbers \citep{Sutter:2012,Paz_2013,Krause:2013,Clampitt:2015,Nadathur:2016}.

We illustrate all our arguments using measurements in simulations which are described in Section~\ref{sec:sims}.  Section~\ref{sec:evolve} shows the short-comings of the spherical evolution model, and Section~\ref{sec:move} describes our modification, and its effect on density and velocity profiles.  Section~\ref{sec:ESTmodel} describes the excursion set troughs model for void profiles, and compares it with our measurements.  A final section summarizes.  Details associated with the excursion set troughs approach are provided in an Appendix.  

\section{Voids in simulations}\label{sec:sims}
We use the voids identified in N-body simulations to illustrate a number of the points raised in the discussion which follows.

The simulations and void catalogs are from \cite{Massara:2015}. We use their low resolution simulations run in a $\Lambda$CDM cosmology, averaging over the $10$ available realizations. For each realization, the initial conditions (ICs) were set at redshift $z=99$ and particle positions and velocities at $z=2,1,0.5,0$ are available. Voids were identified using the void finder {\sc VIDE} \citep{VIDE} at redshift $z=0$ and were divided into three groups, depending on their size at that time: $R_{\rm eff} = 14-16, 16-18, 20-25$ $h^{-1}$Mpc.  \cite[See][for more details.]{Massara:2015}. Also see \cite{Nadathur:2015} for discussion of the subtleties associated with such void finders.  

At $z=0$, each void has a number of member particles.  We define the center of each void at redshift $z\geq 0$ as the center of mass of the $z=0$ member particles.  (This is exactly analogous to how one traces halos back in time to the protohalo patches from which they formed.)  We use the term `protovoid' to refer to the set of void particles at any earlier time (e.g. at $z=99$).  We then compute the cross-correlation between the void or protovoid centers and the matter distribution.  Of course, we are free to cross-correlate the (proto)void centers at one redshift with the matter distribution at another.  If the void centers do not move, then this cross-correlation is a useful measure of how the voids expanded.  However, if the voids moved, then the measurement is more complicated to interpret.

\section{The spherical evolution model}\label{sec:evolve}
The simplest models of halos and voids assume that they form from the spherically symmetric collapse or expansion of sufficiently over or underdense isolated spherical patches in the primordial mass density fluctuation field \citep{gg72, ddglp93}.  In this model, concentric shells remain concentric as they expand (around what will become void centers) or contract (around clusters).
The spherical model predicts a deterministic mapping between the linear overdensity $\delta_{\rm L}(<R_{\rm L})$ within the initial (often called Lagrangian) scale $R_{\rm L}$ and the nonlinearly evolved one $\delta_{\rm E}(<R_{\rm E})$ within the evolved scale $R_{\rm E}$ at redshift $z$.

The key steps in the spherical model follow from the assumption that shells do not cross, so 
\be
 1+\delta_{\rm  E}(<R_{\rm E}|a) = \left[R_{\rm L}/R_{\rm E}(a)\right]^3
                             = 1 + \delta_{\rm Sph}(<R_{\rm E}|a),
 \label{rLrE}
\ee 
where $R_{\rm L}$ is the initial comoving radius of the patch that has comoving size $R_{\rm E}$ at the epoch $a$, and 
\begin{equation}
 1 + \delta_{\rm Sph}(<R_{\rm E}|a) \approx \left(1 - \frac{D_a\delta_{\rm L}(<R_{\rm L})}{\delta_c}\right)^{-\delta_c}
 \label{dEdL}
\end{equation}
\citep{b94,rks98}, where $D_a$ is the linear growth factor normalized to the present time, and $\delta_c=1.686$.  (The spherical model mapping between $\delta_{\rm L}$ and $\delta_{\rm E}$ can be written exactly, but the solutions for over and under-densities appear rather different.  Equation~\ref{dEdL} is a simple but accurate approximation which is valid in all cases.)

Although recent work has shown how to incorporate effects which go beyond the spherical model \citep{scs13}, these only enter at second order in the $\delta_{\rm E}-\delta_{\rm L}$ mapping, so we will ignore them in what follows.  Together, equations~(\ref{rLrE}) and~(\ref{dEdL}) allow one to transform from an initial profile shape to a final one.  

\begin{figure}
 \centering
 \includegraphics[width=0.9\hsize]{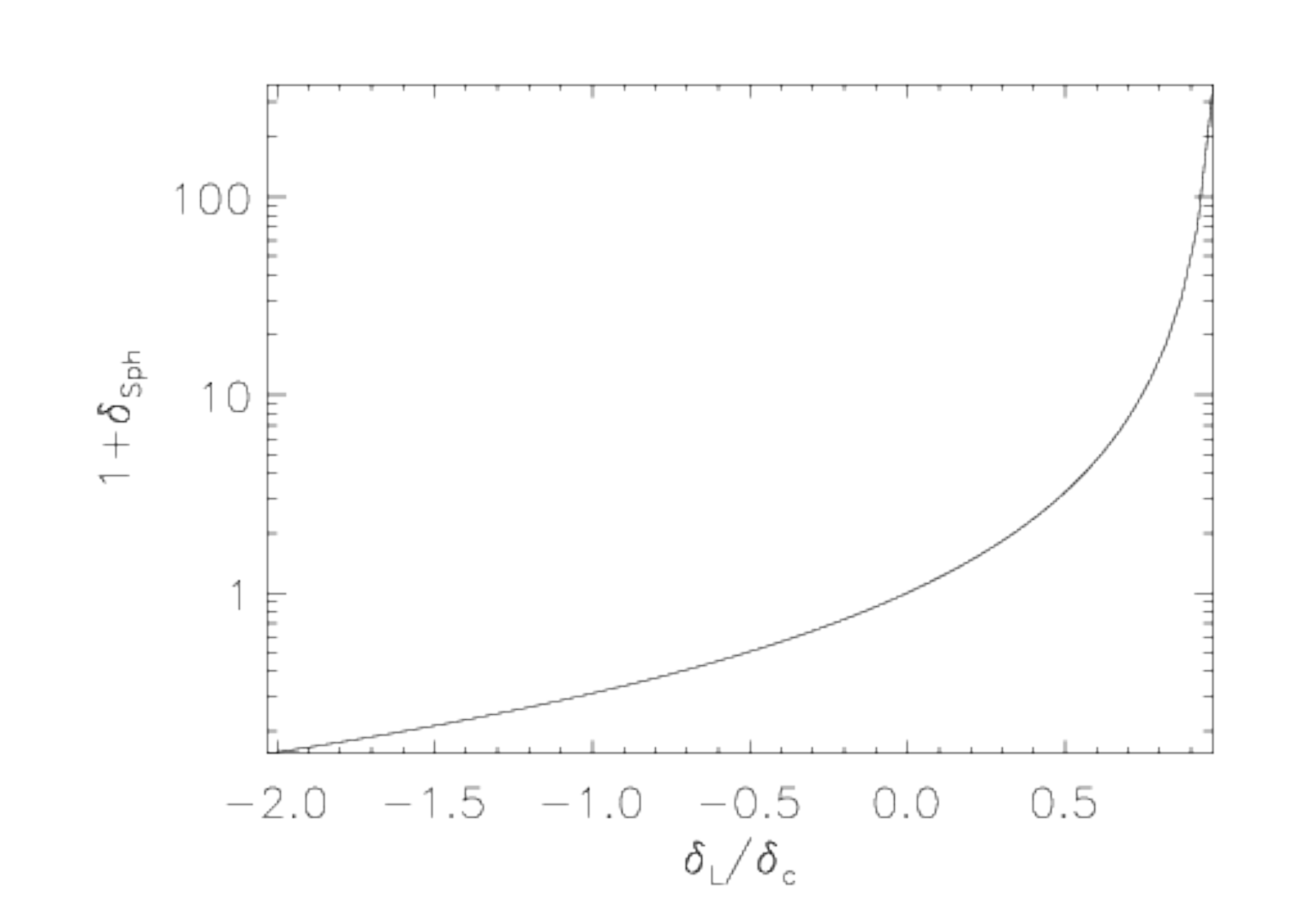}
 \caption{Monotonic mapping between linear and nonlinear density in the spherical model (equation~\ref{dEdL}).  The logarithmic scale on the y-axis shows that a wide range of nonlinear densities (greater than $\sim 100$) all have $\delta_{\rm L}/\delta_c\to 1$, whereas small nonlinear densities (less than $\sim 0.2$) correspond to a rather large range of $\delta_{\rm L}$.  }
 \label{fig:SC}
\end{figure}

Before we show the results of doing so, it is worth making the following point about this mapping.  The Eulerian density diverges rapidly as $D_a\delta_{\rm L}\to\delta_c$ (see Figure~\ref{fig:SC}).  As a result, large changes in $\delta_{\rm E}$ correspond to small changes in $D_a\delta_{\rm L}$. Therefore, the precise value of $\delta_{\rm E}$ that is used to define a halo is not so important for modeling the protohalo patches from which the halo formed.  In contrast, $D_a\delta_{\rm L}\to\infty$ as $1+\delta_{\rm E}\to 0$.  In this limit, which is the relevant limit for voids, small changes in $\delta_{\rm E}$ correspond to large changes in the associated $D_a\delta_{\rm L}$.  As a result, predictions of void formation are rather sensitive to the details of the void-finding algorithm.  

To illustrate, models of the largest halos and voids predict comoving number densities which scale approximately $\propto \exp(-\delta_c^2/2s_0)$ and $\propto \exp(-\delta_v^2/2s_0)$.  Whereas $\delta_c\approx 1.686$ so long as $\delta_{\rm E}$ is more than a few hundred times the background density, $\delta_v$ can depend strongly on the void finder.  The spherical collapse of a tophat profile suggests that the value $D_a\delta_{\rm L} = -2.7$ is special; this value in equation~(\ref{dEdL}) indicates that the associated enclosed density is $\delta_{\rm E}=-0.8$.  However, while the enclosed density asssociated with most void finders is close to this value in the central regions, the value within what is commonly quoted as the `effective radius' of the void is more like $-0.5$.  This corresponds to $D_a\delta_{\rm L}\approx -0.8.$  Such `voids' would be much more abundant than predicted if one assumed $\delta_v=-2.7$.  Accounting for this goes a long way towards explaining many of the reported discrepancies between predicted and measured abundances of voids.  E.g., Chan et al. (2014) report that void abundances in their simulations are reasonably well-predicted if one sets $\delta_v\approx -1$.

\subsection{From Lagrangian to Eulerian}
\begin{figure}
 \centering
 \includegraphics[width=0.9\hsize]{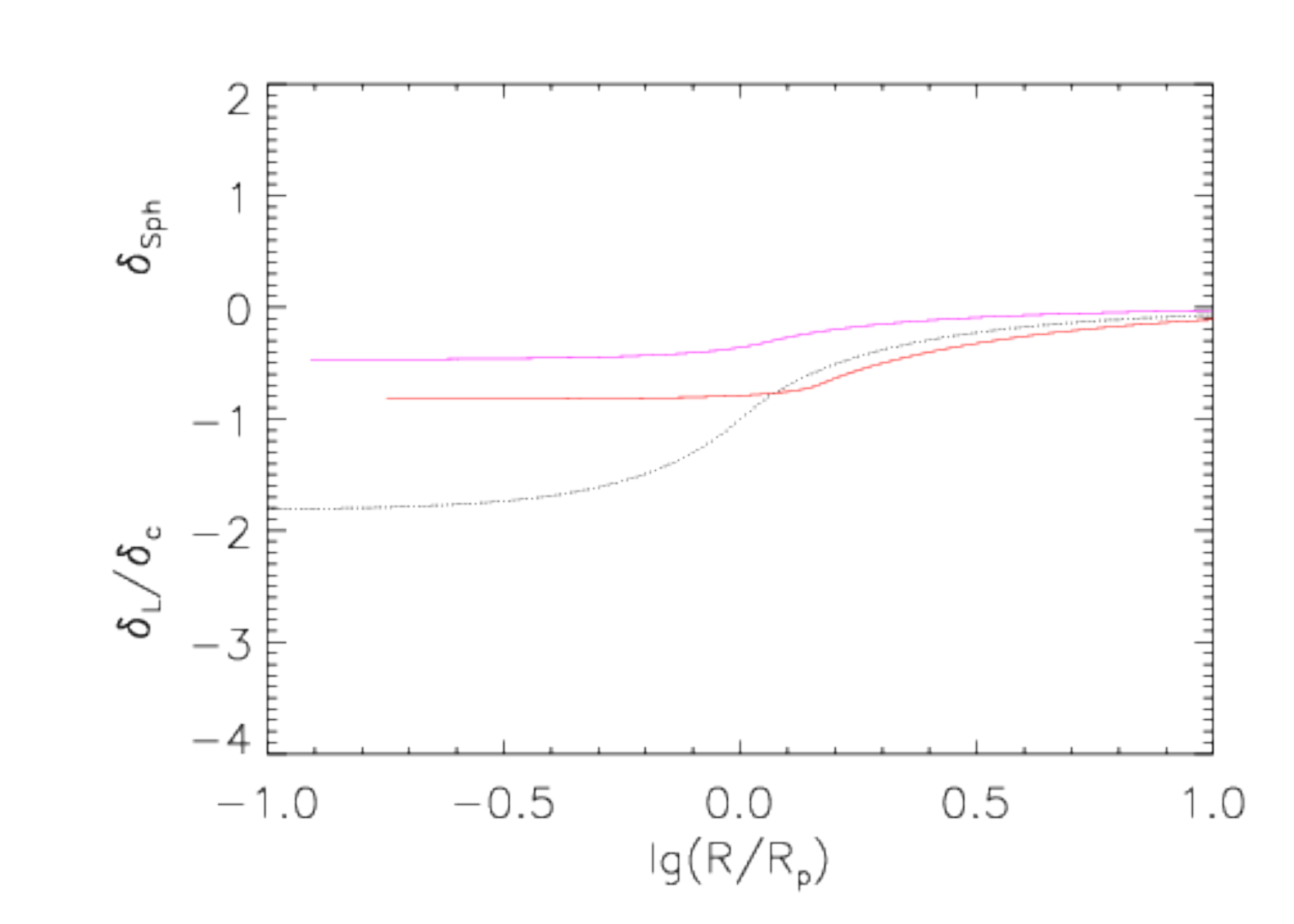}
 \includegraphics[width=0.9\hsize]{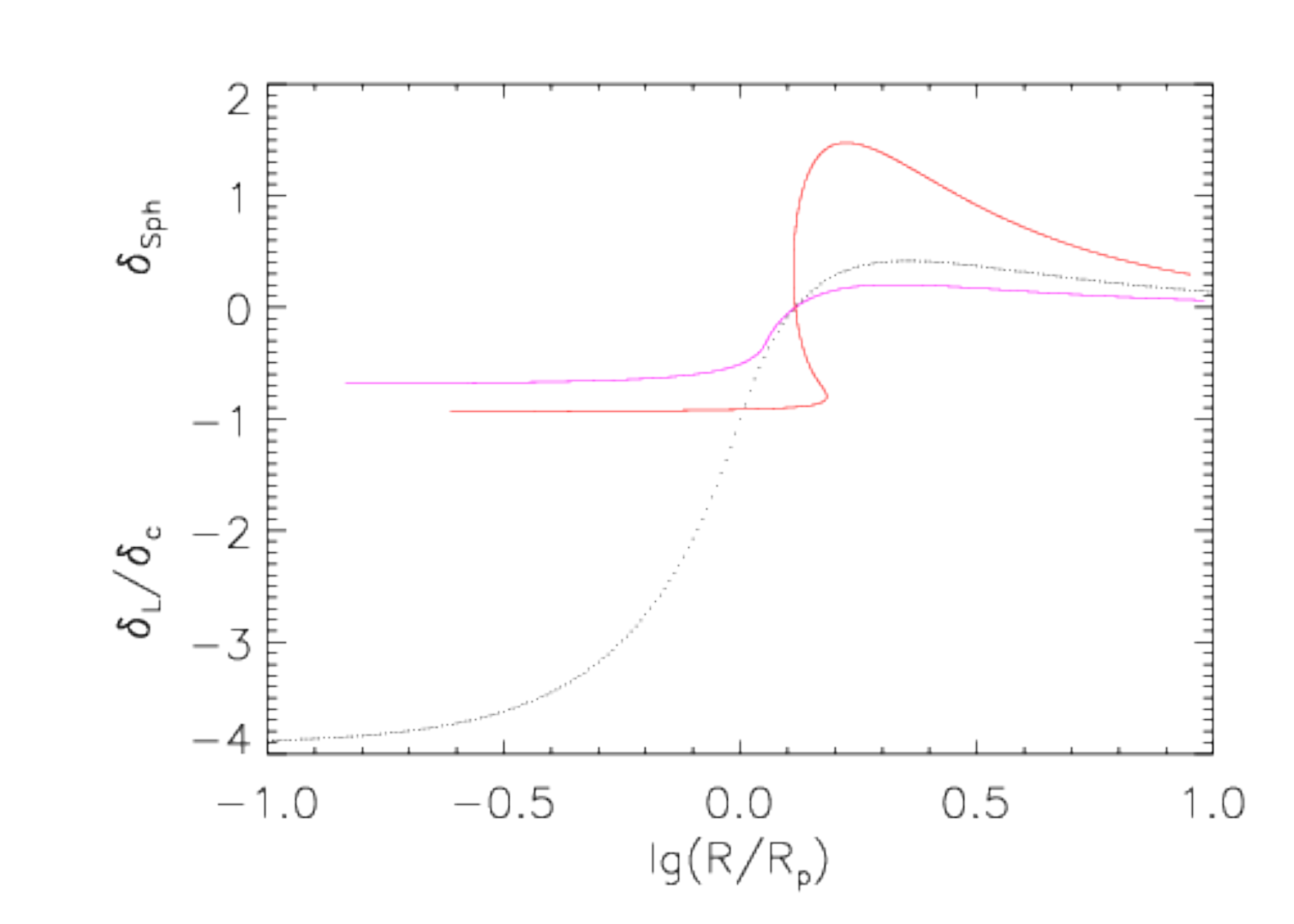}
 \caption{Evolution of the density profile around an excursion set trough of height $\delta_p = -1.686$ defined using TopHat smoothing, when the underlying Gaussian field has $P(k)\propto k^{-2}$.  Dotted curve shows the Lagrangian profile (i.e. the initial one, evolved using linear theory to the present time); solid curves show the nonlinearly evolved profiles (solid) when the linear theory growth factor is $D_a=0.25$ (magenta) and $D_a=1$ (red) times that of the present time.  Top and bottom panels show troughs of size $\nu\equiv \delta_c/\sqrt{s_0^{pp}} = 1$ and 0.25, and were chosen to illustrate the simple case of pure spherical expansion (top), and one where the void-in-cloud process occurs, so the evolution shown is incorrect (bottom).
 }
 \label{fig:dvoidTH}
\end{figure}

Figure~\ref{fig:dvoidTH} illustrates how equations~(\ref{rLrE}) and~(\ref{dEdL}) can be used to map an initial Lagrangian profile to an evolved profile at a later time.  The top panel shows the evolution of the enclosed density profile that is motivated by the excursion set peaks approach which we describe in more detail in the Appendix.  For the current discussion, the main point is simply that, given a value of $D_a$, equation~\ref{dEdL} transforms the dashed curve into the solid ones.  Although the exact shape of the dashed curve differs in detail, the evolution as $D$ increases is similar to that shown in the right hand panel of Figure~3 of \cite{svdw04}:  as time progresses, underdense regions expand and empty-out.

The bottom panel shows the evolution of a Lagrangian enclosed density profile that was initially slightly overdense on large scales (dashed curve is greater than zero).  Within the region where the Lagrangian profile was negative, the Eulerian profile expands outwards and empties out.  However, where the profile was positive, the Eulerian profile shrinks inwards and grows denser.  The zero-crossing scale does not evolve, as can be seen directly by setting $\delta_{\rm }=0$ in equation~(\ref{dEdL}).  As a result, there is a range of scales within which the Eulerian profile is triple valued:  these are the `shell-crossed' scales where the void in cloud process highlighted by \cite{svdw04} is important.  In this regime, the red curves shown are incorrect, and a procedure like that outlined by \cite{pls12} to account for shell-crossing must be used.  E.g. the correct curves would replace the large $R$ small $\delta_{\rm E}$ parts of all `S'-shaped profiles with vertical lines which drop down from the left-most (smallest $R$) tangent point.

\subsection{From Eulerian to Lagrangian}
Although the profiles in the upper panel were obtained by inserting the dashed curve for $\delta_{\rm L}$ in the right hand side of equation~(\ref{dEdL}), and evaluated for different $D_a$, we could, in fact, have done the opposite.  We could have started from any one of the red curves, solved for the shape of $D_a\delta_{\rm L}$ using
\begin{equation}
  \frac{D_a\delta_{\rm L}(<R_{\rm L})}{\delta_c}
  = 1 - \Bigl[1 + \delta_{\rm Sph}(<R_{\rm E})\Bigr]^{-1/\delta_c},
\end{equation}
and hence, have inferred the shape of the red curve associated with any other value of $D_t/D_a$. (I.e., compute the new $1+\delta(a_t)$ associated with $(D_t/D_a)D_a\delta_{\rm L}(<R_{\rm L})$ and then the new $R_{\rm E}(a_t)$.)

\begin{figure}
 \centering
 \includegraphics[width=0.95\hsize]{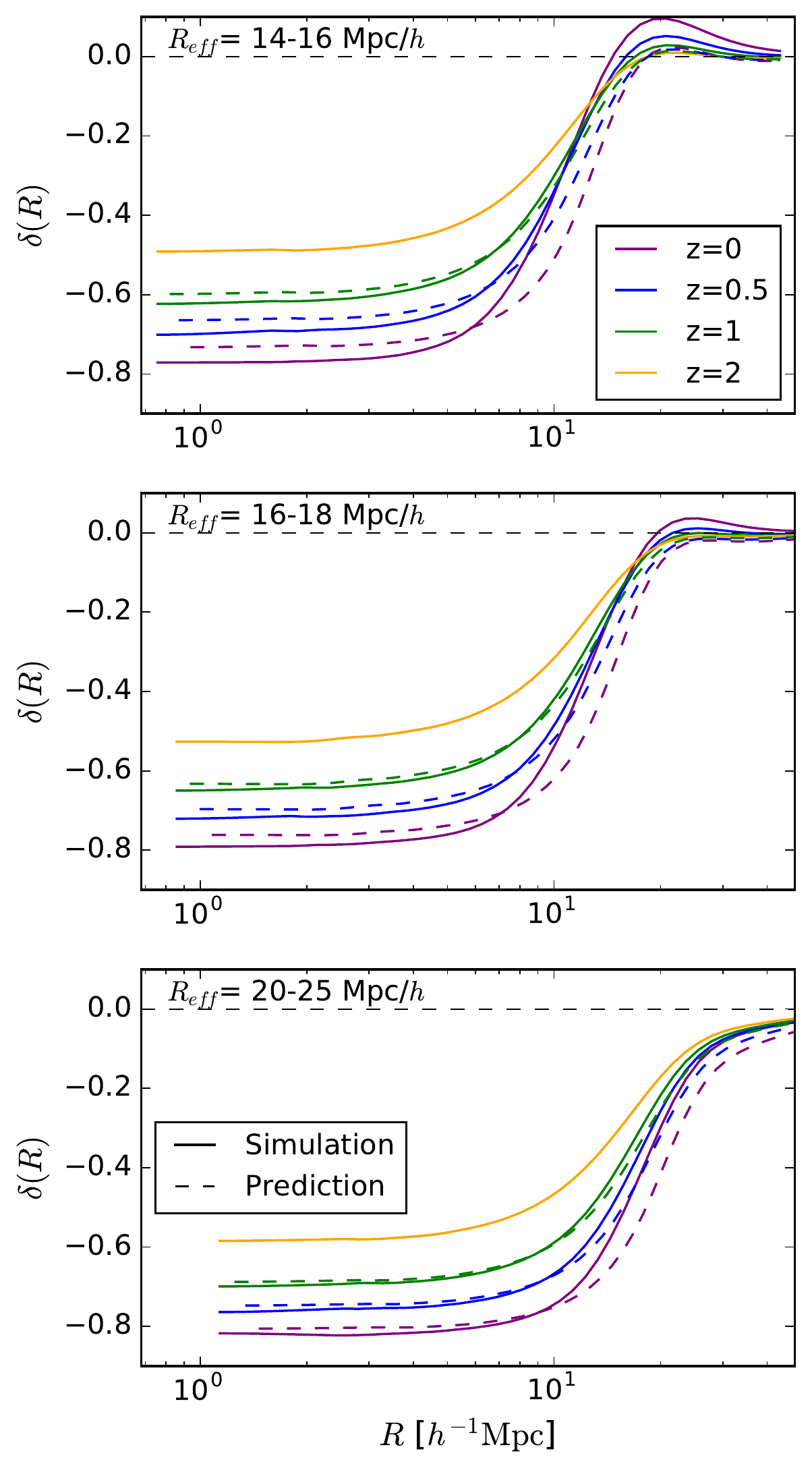}
 \caption{Mean enclosed void density profiles -- i.e. the volume averaged void-matter cross-correlation function -- at $z=2, 1, 0.5$ and 0. Solid lines show the measurement and dashed lines show the result of combining the measured shape at $z=2$ with the spherical model (equation~\ref{dEdL}) to predict the shape at later times.}
 \label{fig:onlySC}
\end{figure}

This is sufficiently straightforward that it is interesting to ask how well this works for voids identified in simulations.  Figure~\ref{fig:onlySC} shows the results.  The solid lines show the evolution of the void-matter cross correlation functions associated with voids that have size $R_{\rm eff}$ (as labelled) at $z=0$, but measured using protovoid centers and the mass distribution both at $z=2, 1$, 0.5 and 0 (top to bottom in each panel).  The dashed curves show the result of using the spherical model to predict the other curves from the upper most solid curve (i.e. to predict the lower redshift profiles from that at $z=2$).  While the predictions are qualitatively accurate, they lie systematically below the measurements on large scales.  This is true even for the largest voids, which do not have overdense walls, so the complications of the void-in-cloud process do not arise.

The discrepancy on the largest scales is particularly worrisome, since such scales are generally thought to be `in the linear regime' and hence, ought to be the easiest to use for constraining cosmological models.  What has gone wrong?

\subsection{Non-commutation between averaging and evolution?}
One possibility is that, because it is a cross-correlation, each solid curve represents an average over many different void profiles.  It is not obvious that the average of the evolved profiles (which the solid lines represent) should equal the evolution of the average profile (which is what our procedure for computing the dashed curves assumes).  Non-commutation between averaging and evolution would be particularly relevant if there were significant scatter around the mean profile.  

\begin{figure}
 \centering
 \includegraphics[width=0.95\hsize]{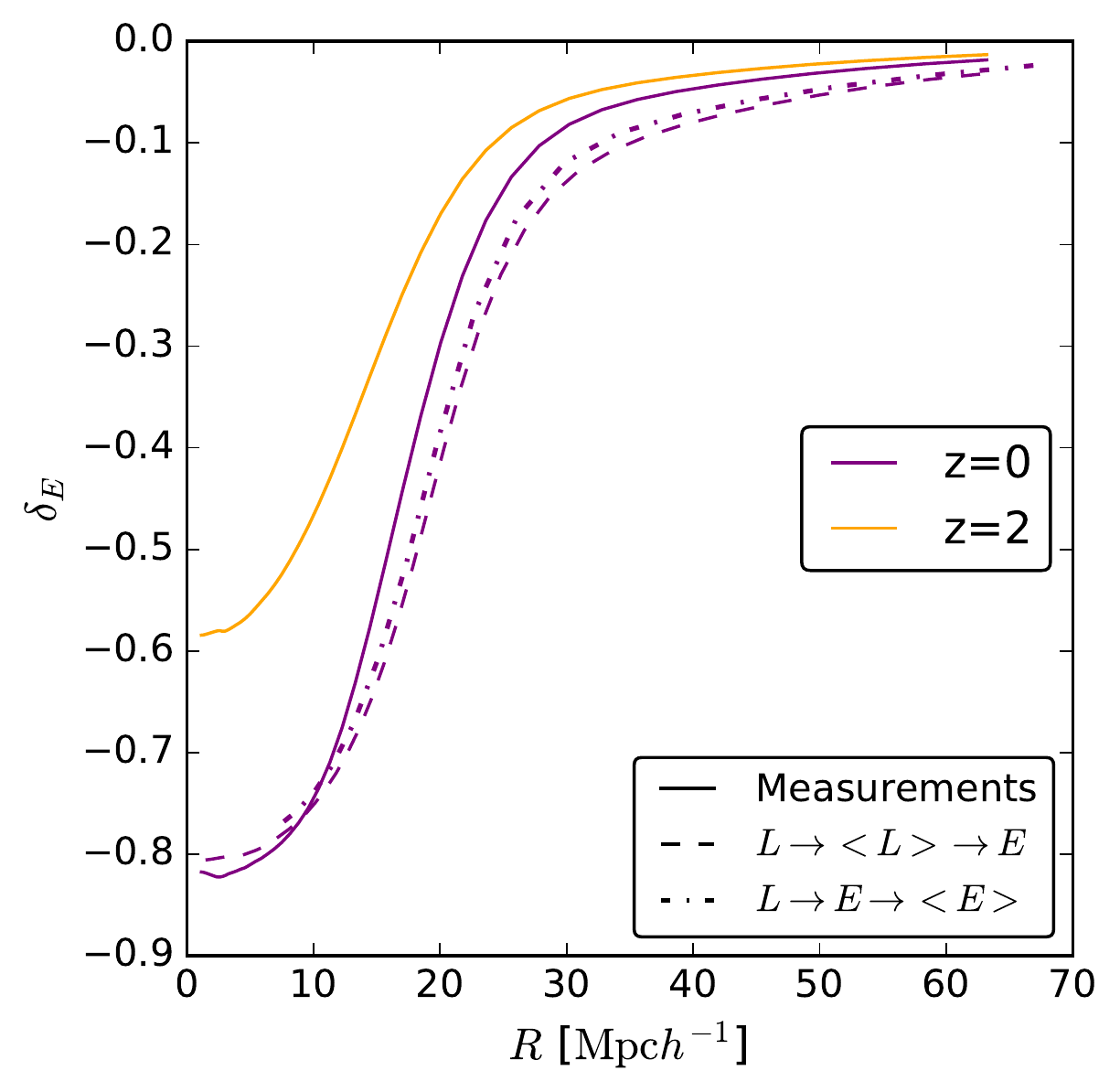}
 \caption{Comparison of profiles obtained from evolving the mean $z=2$ profile (orange solid) to $z=0$ (dashed) with those from averaging the evolved profiles (dot-dashed).  The dashed curves are the same as the dashed curves shown in the bottom panel of the previous Figure:  they show how the (volume averaged) void-matter cross-correlation function measured at $z=2$ is predicted to evolve if the spherical model (equation~\ref{dEdL}) were accurate.  While there are differences on small scales, the two are very similar on large scales.  Both lie below the black solid curve, which shows the actual mean profile at $z=0$.}
 \label{fig:noCommute}
\end{figure}

Figure~\ref{fig:noCommute} provides a direct test.  The solid orange and purple curves are the same as the $z=2$ and $z=0$ profiles shown in the bottom panel of the previous figure.  The dashed curve shows the result of evolving the orange curve to $z=0$ using the spherical model (equations~\ref{rLrE} and~\ref{dEdL}).
I.e., it shows the predicted evolution of the average profile.  The dot-dashed curve shows the average of the evolved profiles:  it was obtained by taking each $z=2$ profile, evolving it to lower $z$ using the same spherical model, and then averaging the result.  The dashed and dot-dashed curves differ on small scales, because evolution and averaging do not commute.  However, on large scales, there is good agreement.  Therefore, although one can account for the scatter in profile shapes and evolution following \cite{rks98}, this will not solve the problem that the predicted evolution does not change the large scale bias.

\begin{figure}
 \centering
 \includegraphics[width=0.95\hsize]{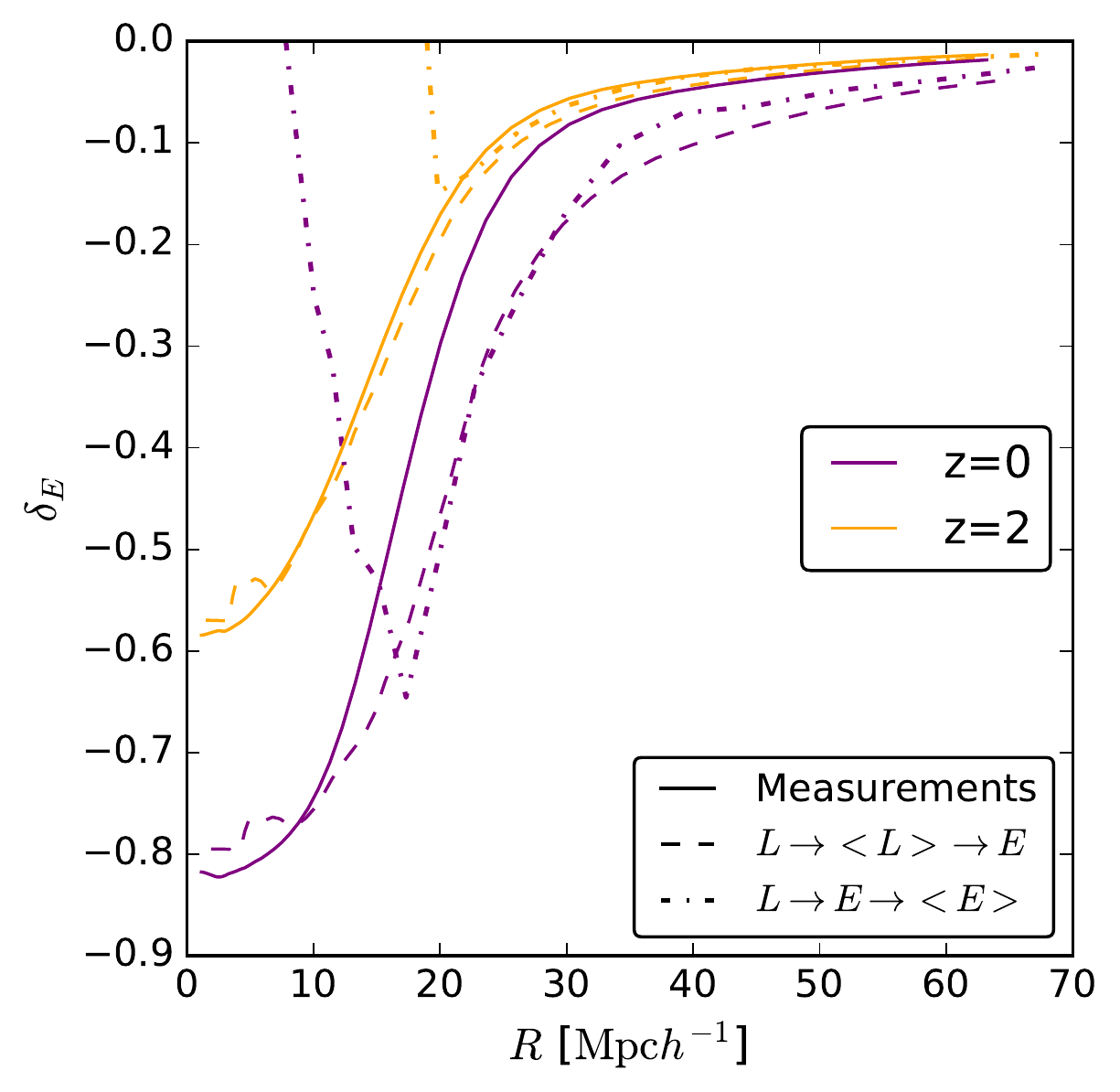}
 \caption{Same as previous figure, but now for evolving the $z=99$ profile to $z=2$ and to $z=0$.  The jump at $R\sim 20h^{-1}$Mpc is due to discreteness effects in the $z=99$ profile.}
 \label{fig:noCommute99}
\end{figure}

One might worry that, because we started with $z=2$ profiles, we have already mixed-up some of the evolution and averaging.  (This is true, of course, but it would not be an issue if the two commute.)  Therefore, Figure~\ref{fig:noCommute99} shows the result of starting with the $z=99$ profiles instead, and then evolving the average profile to $z=2$ and $z=0$ (dashed curves).  The dot-dashed curves evolve and then average.  The upturn on small scales is due to discreteness effects in the simulations (which is why our prefered default is to work with the $z=2$ profiles).  This upturn does not complicate the main point, which is that neither the dashed nor the dot-dashed curves look like the mean profile measured in the simulations (solid curves).  Whereas the differences are not large at $z=2$, they are significantly larger at $z=0$.  

Together, Figures~\ref{fig:noCommute} and~\ref{fig:noCommute99} show that the real fault lies not with the fact that evolution and averaging do not commute, but with equation~(\ref{dEdL}) for the evolution itself.  We address this in the next section.

\section{Spherical evolution around a moving center}\label{sec:move}

\subsection{Large-scale bias}
As \cite{svdw04} note, equation~(\ref{dEdL}) is supposed to describe the evolution of the profile of an {\em isolated} object.  Since all objects are embedded in the cosmic web,  equation~(\ref{dEdL}) cannot be the full story.  To see why, first note that the procedure used for measuring a `void profile' is the same as that for measuring the cross-correlation between void centers and the dark matter distribution.  Therefore, one should think of the two as equivalent:
\begin{equation}
  1 + \bar\xi_{\rm bm}^{\rm E}(x|a) \equiv
  1 + \frac{3}{r^3}\int_0^r {\rm d}x\,x^2 \xi_{\rm bm}^{\rm E}(x|a) =
  1 + \delta_{\rm E}(<r|a). 
  \label{dExibar}
\end{equation}
This is instructive because it has been known for decades that, if the initial conditions were Gaussian, then the large scale cross-correlation function will have the same shape as the dark-matter auto-correlation function, although it may differ in amplitude:
\begin{equation}
 \delta_{\rm E}(<r|a)\approx b_{\rm E}(a) \, \bar\xi_{\rm mm}(<r|a)
 \label{linearb}
\end{equation}
\citep{bbks86}.
The constant of proportionality is known as the `linear bias factor'.
We note in passing that, therefore, `universal' fitting formulae for the `void profile' shape which do not exhibit this property (sufficiently far from the void center) are not `universal'.  E.g., in this sense, the formula of \cite{HSW_2014} is not universal \citep[right-most panel of Figure~6 in][]{chd14}: it is only useful on smaller scales (although the scale beyond which it is inaccurate is not well-defined), and it is not very useful if, as is often the case, one is interested in its Fourier transform.  

\subsection{Evolution of bias}
Next, note that if the comoving number density of biased tracers is conserved, then the bias factors at two different times are related:  
\begin{equation}
 b_{\rm E}(a_1) - 1 = (D_2/D_1)\,(b_{\rm E}(a_2) - 1).
 \label{bconserved}
\end{equation}
\citep{ck89, nd94}.
It is conventional to define
\begin{equation}
 b_{\rm L} \equiv b_{\rm E}(a=1) - 1
\end{equation}
\citep{mw96},
so that 
\begin{equation}
 b_{\rm E}(a) = (D_a + b_{\rm L})/D_a.
 \label{bEbL}
\end{equation}
This is a rather generic argument which implies that the amplitudes of the Lagrangian and Eulerian cross-correlation functions should be different.  In contrast, simply using equation~(\ref{dEdL}) to evolve the Lagrangian profile into an Eulerian one will result in no change to the profile (shape or amplitude) on the large scales where $\delta_{\rm L}\ll 1$.  Namely, to leading order in $\delta_{\rm L}$, the spherical evolution equation~(\ref{dEdL}) has 
\begin{equation}
  \delta_{\rm E}(<R_{\rm E}|a) \approx D_a\,\delta_{\rm L}(<R_{\rm L})
  \approx D_a\,b_{\rm L}\, \bar\xi_{\rm mm}(<r|a=1).
  \label{bnomotion}
\end{equation}
Compared to equation~(\ref{linearb}) with equation~(\ref{bEbL}), this is missing a factor of $D_a^2\bar\xi_{\rm mm}$.

In suggestive notation, if we define $\delta_0\equiv \delta_m(a_i)/D_i$ and $\delta_b(a)\equiv b_{\rm E}(a)\,\delta_m(a)$ then
\begin{align}
 \langle\delta_b(a)\delta_m(a)\rangle
   &= b_{\rm E}(a)\,\langle\delta_m(a)\delta_m(a)\rangle \nonumber\\
   &\approx (D_a + b_{\rm L})\,\langle\delta_0\,\delta_m(a)\rangle,
\end{align}
where we have approximated $\delta_m(a)/D_a \approx \delta_m(a_i)/D_i$.  This expresses the cross-correlation between biased tracers (e.g. halos or voids) and the matter field at the same epoch $a$, and should be reasonably accurate on large scales.  Cross-correlating the positions of the biased tracers at $a=1$ with the mass at any other $a\le 1$ yields 
\begin{equation}
 \langle\delta_b(a=1)\delta_m(a)\rangle
   \approx (1+b_{\rm L})\,\langle\delta_0\,\delta_m(a)\rangle
\end{equation}
since $D_a=1$ when $a=1$.
Alternatively, cross-correlating the protohalo or protovoid positions at $a_i\ll 1$ with the mass at any $a$ yields 
\begin{equation}
 \langle\delta_b(a_i)\delta_m(a)\rangle
   \approx b_{\rm L}\,\langle\delta_0\,\delta_m(a)\rangle,
\end{equation}
where we have assumed $D_i\to 0$. Thus, the Lagrangian and Eulerian density profiles (i.e., the profiles around the biased positions at $a_i$ and at $a=1$) are expected to be different, even on large scales.

\begin{figure}
 \centering
 \includegraphics[width=0.9\hsize]{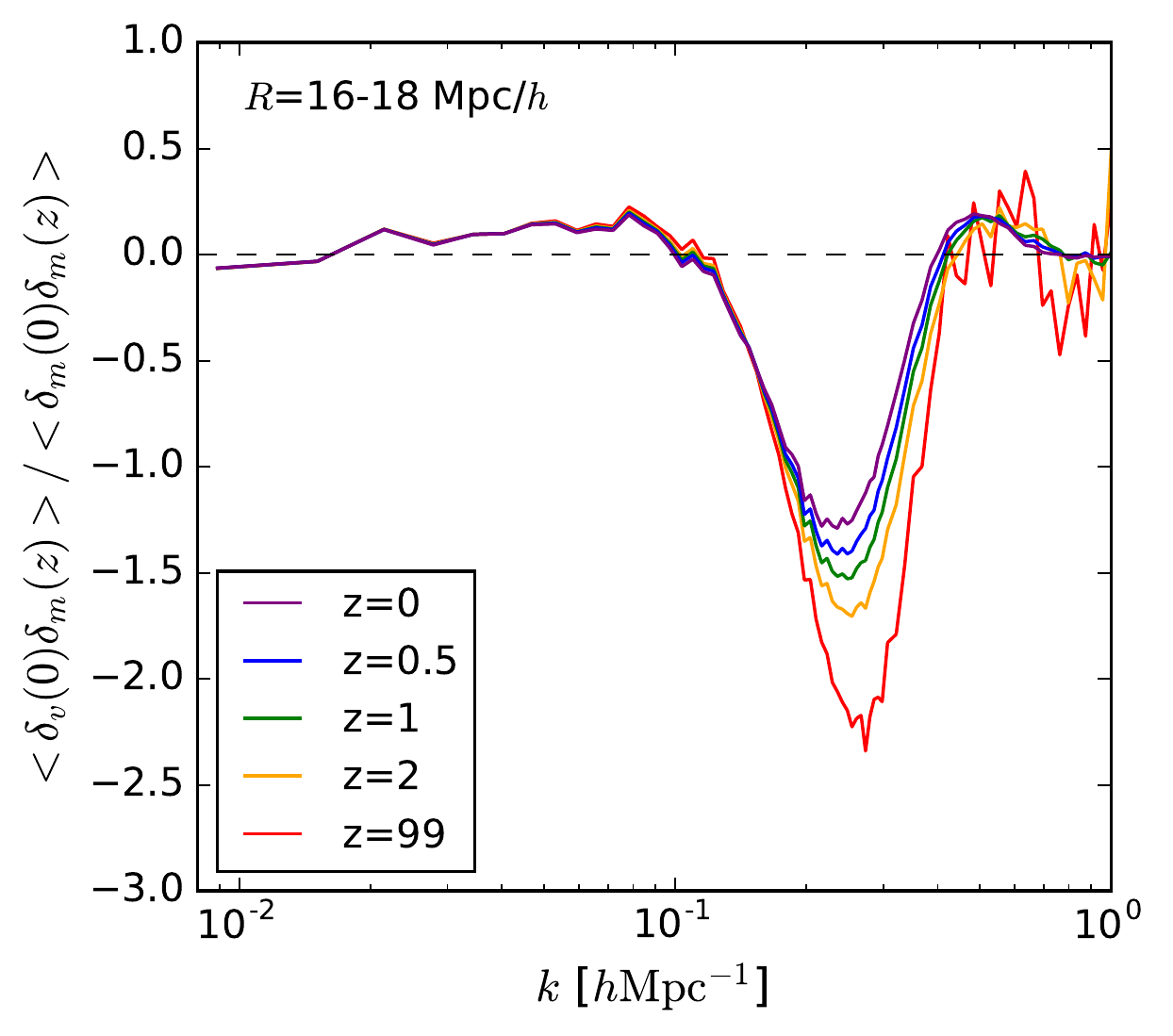}
 \caption{The void matter-cross power spectrum -- i.e. the Fourier transform of the density profile around positions identified as void centers at $z=0$.  At small $k$ (i.e., on large scales) the profile around these positions is the same at all redshifts.}
 \label{b0}
\end{figure}

\begin{figure}
 \centering
\includegraphics[width=0.9\hsize]{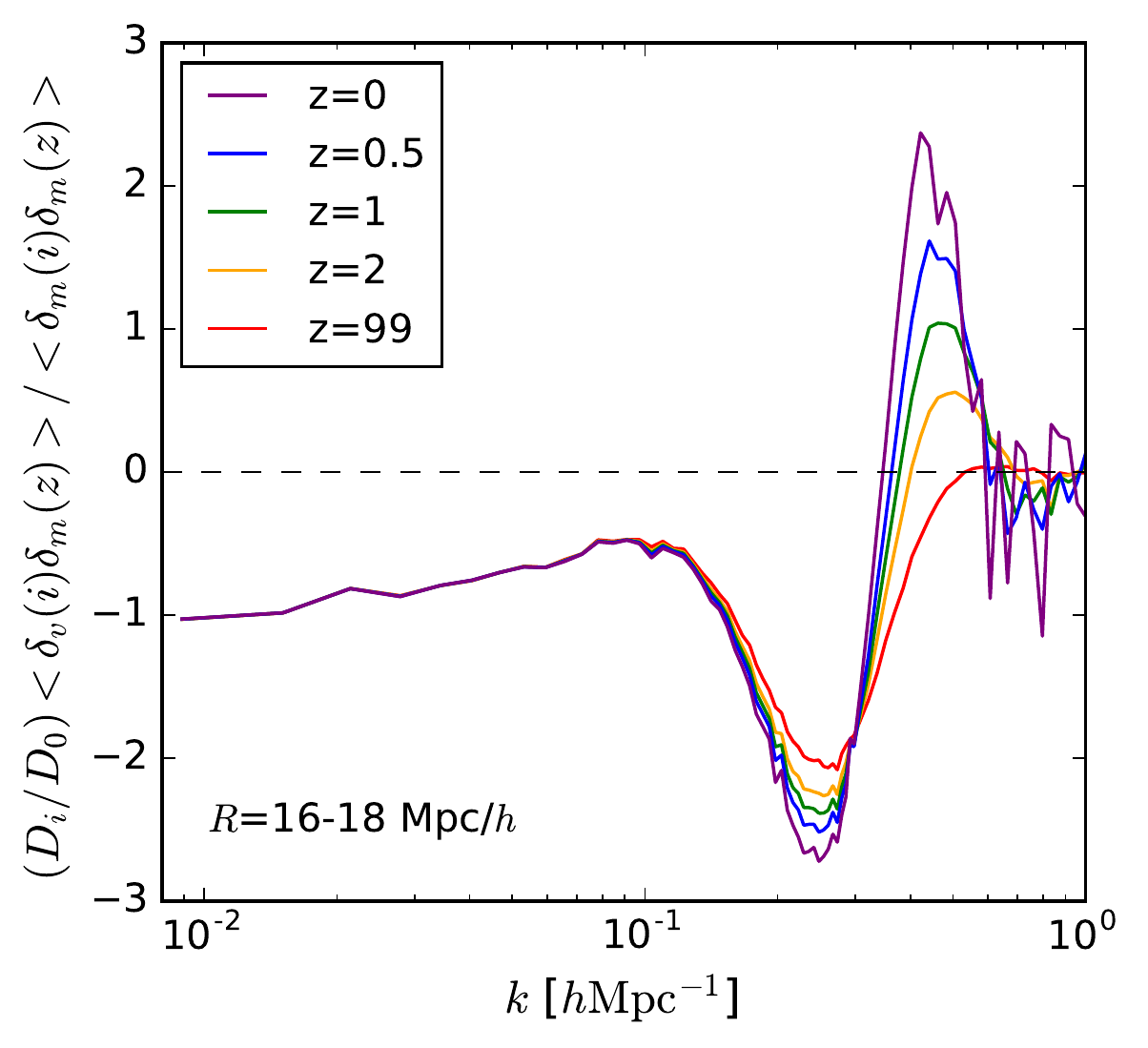}
\caption{Same as previous figure, but now showing the evolution of the matter distribution around the protovoid centers in the initial conditions ($z=99$).  Again, there is no evolution at small $k$.  However, the small $k$ limit differs by $-1$ from that in the previous figure.}  
\label{b99}
\end{figure}

Figure~\ref{b0} shows the Fourier transform of the void-matter cross-correlation function using the $z=0$ void centers and the matter distribution at other redshifts.  Although there are large differences at large $k$, there is no evolution at small $k$.  Figure~\ref{b99} shows a similar study of the evolution around the protovoid centers at $z=99$.  Again, there is no evolution at small $k$.  However, the $k\to 0$ limit, the large scale bias factor, differs from that in the previous figure:  the difference between the two bias factors equals unity, in agreement with equation~(\ref{bEbL}).  

The large scale amplitude and detailed scale dependence are different if one considers larger or smaller voids (larger voids have more negative bias), but the fact that large scale bias does not evolve if the void center is fixed to its position at one epoch, is generic.  And, if one uses the $z$-dependent protovoid center when cross-correlating with the matter distribution at $z$, then equation~(\ref{bEbL}) provides an excellent description of the large scale bias.

\subsection{Evolution and motion}
In the context of peak theory, equation~(\ref{bEbL}) is a consequence of the displacement of proto-peaks (or troughs) from their initial positions \citep{dcss10}.  Massara \& Sheth (in prep) show that, if the biased tracers do not move, then the same reasoning as in \cite{dcss10} leads to $b_{\rm E}(a) = b_{\rm L}/D_a$ rather than equation~(\ref{bEbL}).  This makes $\langle\delta_b(a)\delta_m(a)\rangle\approx b_{\rm L}\,\langle\delta_0\,\delta_m(a)\rangle$ whether one centers on positions at $a_i$ or $a=1$, since, because they do not move, $\delta_b(a) = b_{\rm L}\,\delta_m(a)/D_a = b_{\rm L}\,\delta_0$ is independent of $a$.  

\begin{figure}
 \centering
 \includegraphics[width=0.95\hsize]{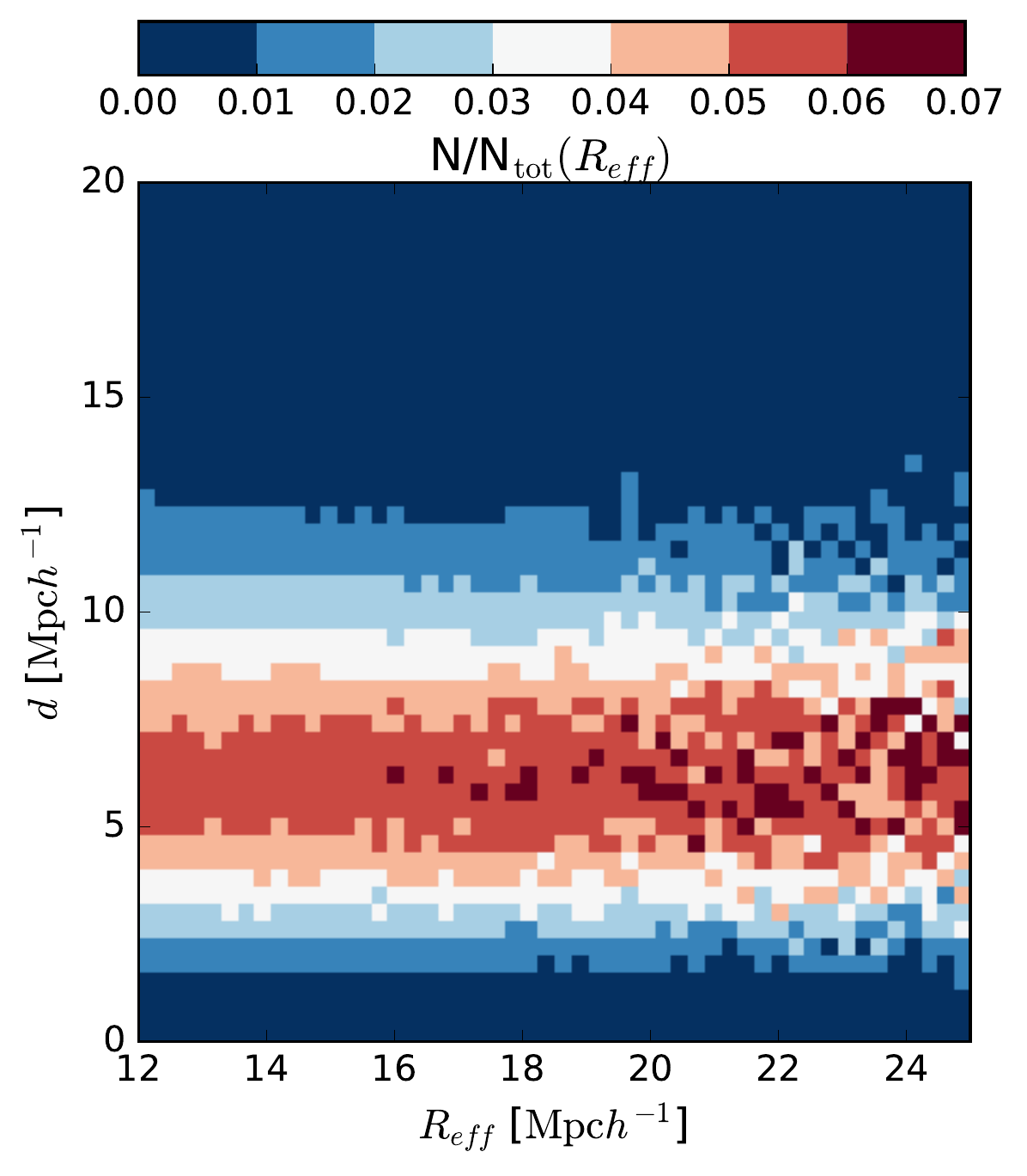}
 \caption{Displacement $d$ between the void center at $z=0$ and the center of the protovoid patch from which it formed, as a function of void size $R_{\rm eff}$.  The color coding is proportional to the fraction of voids of a given size which have displacement $d$, and shows that the distribution of $d$ is the same for all $R_{\rm eff}$.}
 \label{fig:displacements}
\end{figure}

The fact that the $k\to 0$ limit in Figure~\ref{b0} differs from that in Figure~\ref{b99} by unity indicates that voids move.  Figure~\ref{fig:displacements} shows a more direct measure of this motion:  the distance between the void and protovoid centers, plotted as a function of void size.  Voids at $z=0$ are typically about $6h^{-1}$Mpc away from where they started, but this displacement is approximately independent of void size.  Since this displacement is a smaller fraction of the size of a larger void, one might have thought that neglecting void motion would be an acceptable approximation for the larger voids.  While this is true, we have shown that, in fact, this motion has a significant impact on the void-matter cross-correlation in Fourier space.  The only sense in which the impact is smaller for large voids is that they tend to have $|b_{\rm L}|\gg 1$ so, for large voids, $b_{\rm L}+1\approx b_{\rm L}$.  

In summary:  Using equation~(\ref{dEdL}) to model nonlinear evolution leads to the prediction that the large scale bias does not evolve.  Non-evolving large-scale bias is only correct if the center of collapse (or expansion) is not moving.  If it does move, then the Eulerian and Lagrangian bias factors should differ by unity.

\subsection{An improved model}
Having motivated why equation~(\ref{dEdL}) does not model the evolution of the void-matter cross-correlation function, we now test the predictions of a simple extension which accounts crudely but effectively for void motions.  Namely, we consider 
\begin{equation}
  1 + \delta_{\rm E}(<R_{\rm E}|a) = 1 + \delta_{\rm Sph}(<R_{\rm E}|a)
                                    + D_a\,\delta_2(R_{\rm E}|a)
 \label{dEdLcombined}
\end{equation}
where $\delta_{\rm Sph}$ is given by equation~(\ref{dEdL}) and 
\begin{equation}
  \delta_2(R_{\rm E}|a) = D_a\int \frac{{\rm d}k\,k^2}{2\pi^2}\,P_{\rm L}(k)\,
                        b_{\rm v}(kR_p)\, W(kR_p)\,W_{\rm TH}(kR_{\rm E}) 
  \label{d2}
\end{equation}
with 
\begin{equation}
  W_{\rm TH}(x) = 3\, (\sin x - x\,\cos x)/x^3 ,
  \label{Wth}
\end{equation}
\begin{equation}
  W(x) = W_{\rm TH}(x)\,{\rm e}^{-(x/5.5)^2/2} , 
  \label{Weff}
\end{equation}
and
\begin{equation}
  b_{\rm v}(kR_p) = 1 - k^2 \,s_0^{pp}/s_1^{pp}
  \label{bvel}
\end{equation}
with
\begin{equation}
  s_j^{pq} = \int {\rm d}k\,\frac{k^2P(k)}{2\pi^2}\, W(kR_p)\,W_q(kR_q)\, k^{2j},
  \label{sjpq}
\end{equation}
where $W_q = W(kR_q)$ when $R_q=R_p$ but $W_q = W_{\rm TH}(kR_q)$ otherwise.

The addition of $\delta_2$ is a crude way of accounting for the correlated pairs which would be present even if $b_{\rm L} = 0$.  Its form is motivated by the excursion set peaks approach which we describe in Appendix~A.  Briefly, the smoothing windows are related to the void size and to the fact that we are measuring the cross-correlation on scale $R_{\rm E}$.  Although (in the context of the spherical evolution model) a TopHat filter is the most natural choice, we have included an exponential factor in equation~(\ref{Weff}) which softens the edges of the TopHat.  This is motivated by the work of \cite{css17w}.

In addition to the the smoothing windows, $\delta_2$ also depends on $b_v$ (equation~\ref{bvel}).  First, note that if we set $b_v\to 0$, then $\delta_2\to 0$, and the model reduces to the original model of isolated spherical evolution.  The $k$-dependence of $b_v$ is motivated by the excursion set peaks/troughs approach, in which peak/trough motions are expected to differ from that of the dark matter \citep{bbks86,sd01,ds10,dcss10}.  Notice that $\delta_2(R_p)\approx 0$ since most of the contribution to the integral comes from where the difference between $W$ and $W_{\rm TH}$ is small.

We have also explored replacing 
\begin{equation}
  D_a\,\delta_2(R_{\rm E})
  \to \Bigl[1 - D_a\delta_2(R_{\rm L})/\delta_c\Bigr]^{-\delta_c} - 1,
\end{equation}
or setting $\delta_2=0$, and replacing
\begin{equation}
 D_a\,\delta_{\rm L}\to D_a\,(D_ab_v + b_{\rm L})\,\delta_{\rm L}.
\end{equation}
The former tries to write the additional pairs as perturbations to linear theory, whereas the latter subsumes everything into a single spherical collapse like term.  To lowest order, all three models for $1+\delta_{\rm E}$ are the same, so in what follows, we stick with the simplest case (equation~\ref{dEdLcombined}).  
 
\begin{figure}
 \centering
 \includegraphics[width=0.95\hsize]{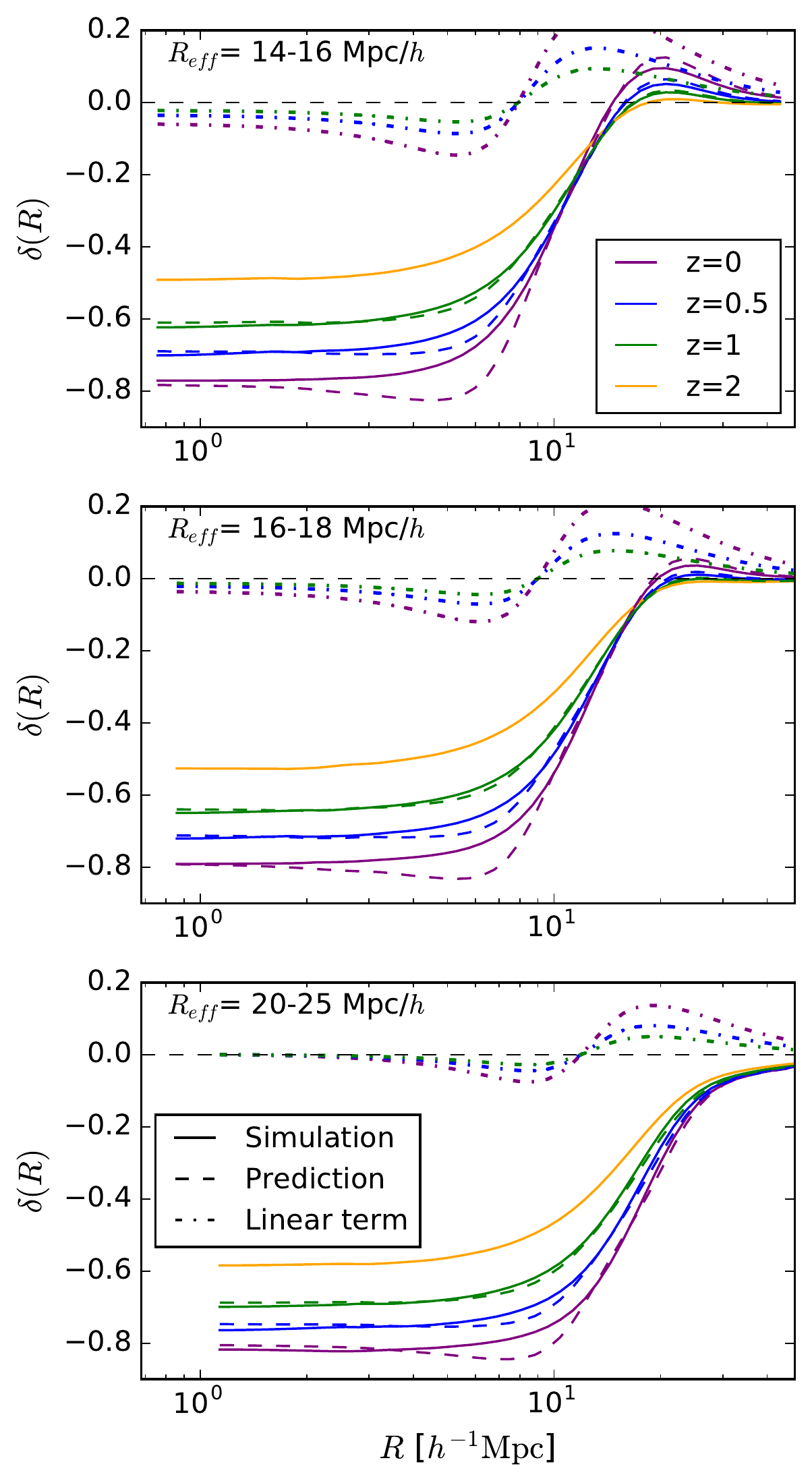}
 \caption{Mean enclosed void density profiles -- i.e. the volume averaged void-matter cross-correlation function -- at $z=2, 1, 0.5$ and 0.  Solid lines show the measurement (same as Figure~\ref{fig:onlySC}) and dashed lines show the prediction from equation~(\ref{dEdLcombined}), which combines the spherical model with an additional term which allows for the evolution of large scale bias.  The dot dashed curves show this extra term. }
 \label{fig:complete}
\end{figure}

\subsection{Tests of the modified model}
Figure~\ref{fig:complete} shows that this model (dashed) describes the large scale profiles (solid) much better than the spherical model alone (compare solid and dashed curves in Figure~\ref{fig:onlySC}).  Here, as before, the profile shape at $z=2$ is used to predict the shape at later times.  The only difference is that now we use equation~(\ref{dEdLcombined}) to model the evolution.  Dot-dashed lines show the extra term compared to equation~(\ref{dEdL}).  Except for the few Mpc around the scale where the profile begins to climb upwards, this extra term leads to a more accurate prediction.

\begin{figure}
 \centering
 \includegraphics[width=0.95\hsize]{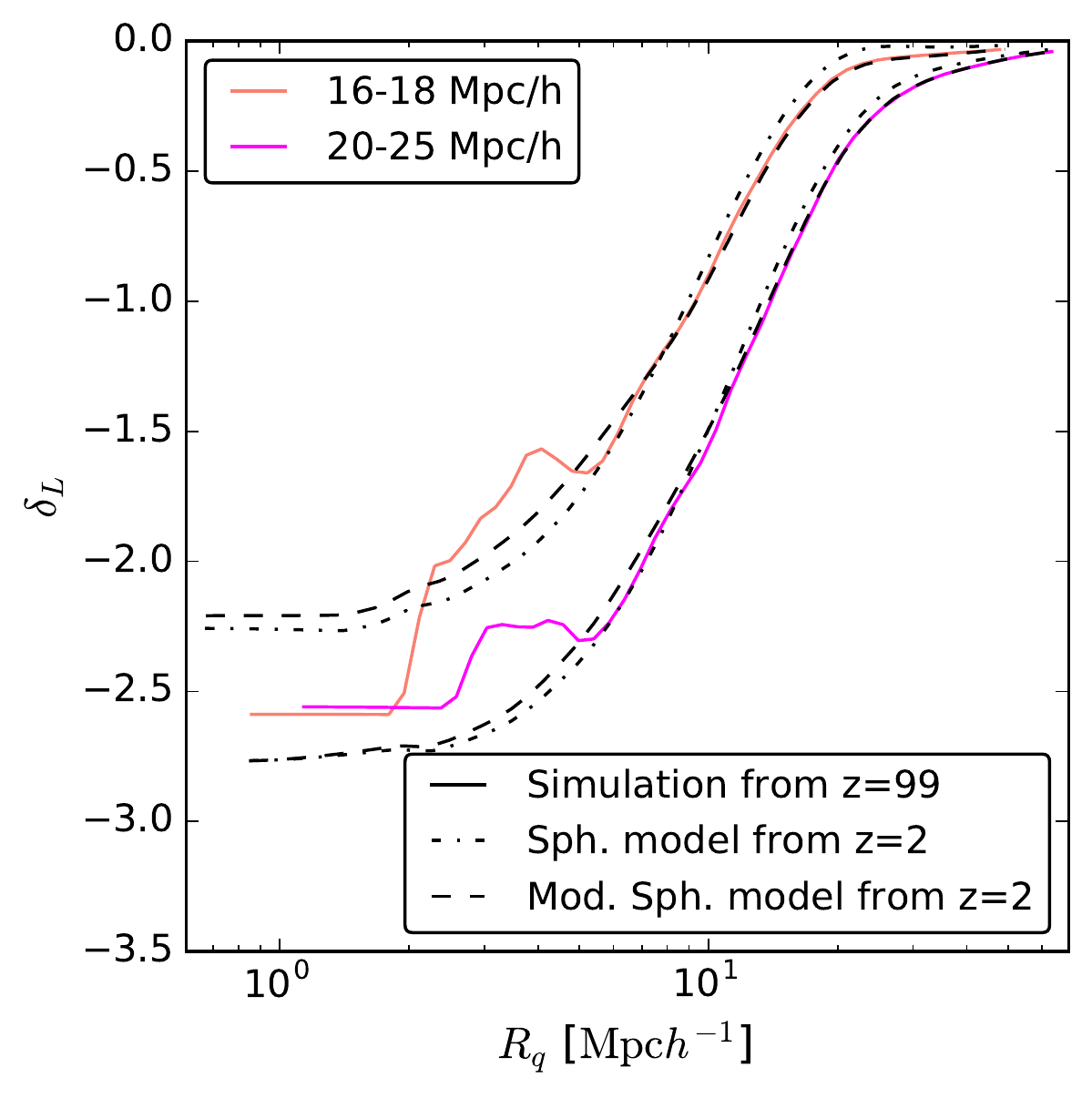}
 \caption{Measured (solid) and predicted Lagrangian profiles.  The predictions result from combining the profile at $z=2$ with equations~(\ref{dEdL}) (dot-dashed) and~(\ref{dEdLcombined}) (dashed) to predict $\delta_{\rm L}$ at $z=99$.  Equation~\ref{dEdLcombined} produces a more accurate description of the measurements.}
 \label{fig:dLag}
\end{figure}

Figure~\ref{fig:dLag} shows an additional test.  The solid curve shows the measured Lagrangian profile $\delta_{\rm L}$, and dot-dashed and dashed curves show the prediction, based on the measured $z=2$ profile and equation~(\ref{dEdL}) and~(\ref{dEdLcombined}), respectively.  The bump in the measurements on small scales is an artifact of the fact that the mean particle separation is $\sim 4h^{-1}$Mpc:  the measurements are only reliable on larger scales.  This bump is the reason why we chose to normalize our model at $z=2$, and then evolve it to lower and higher redshifts, rather than normalizing directly in the initial conditions.  The dashed lines are clearly in better agreeement with the solid ones, indicating that equation~(\ref{dEdLcombined}) is more accurate than equation~(\ref{dEdL}).  

We remarked earlier that, on large scales, our equation~(\ref{dEdLcombined}) has the profile scaling as $D_a (D_a + b_{\rm L})$ (we have set $b_v=1$ in equation~\ref{xi_largeScale}), with $0< D_a<1$.  This shows that the evolution is initially dominated by the piece that is linear in $D_a$, which will have the same sign as $b_{\rm L}$.  At late times, the $D_a^2 > 0$ term will also matter.  Therefore, if $b_{\rm L}<0$, then the profile initially becomes more negative as $D_a$ increases.  The 'turnaround' of the amplitude, due to the $D_a^2$ term, will happen when $d[D_a (D_a + b_{\rm L})]/dD_a = 2D_a + b_{\rm L} = 0$.  I.e., from $D_a > -b_{\rm L}/2$ the amplitude will begin to increase.  However, if $b_{\rm L}$ is too negative, then the critical $D_a$ will be greater than unity; this means that the profile becomes increasingly negative all the way to $z=0$.  This explains why, for the solid curves in Figure~\ref{fig:onlySC}, the late time profile for small voids lies above that at earlier times, whereas the late time profile for big voids becomes ever more negative.

Our evolution model, equation~(\ref{dEdLcombined}), is phrased in a rather generic way.  Except for our choice for $b_v$, it is not tied to a particular functional form for $\delta_{\rm L}$.  Having shown that it is accurate, it is interesting to ask if the shape of $\delta_{\rm L}$ can be predicted from first principles.  We address this in Section~\ref{sec:ESTmodel}.  

\subsection{Velocities}\label{sec:v12}
The pair conservation (or continuity) equation states that
\begin{equation}
 \frac{\partial\bar\xi_{\rm bm}(<r|a)}{\partial\ln a} = -\frac{v_{\rm bm}(r|a)}{Hr}\,3\,[1+\xi_{\rm bm}(r|a)],
 \label{continuity}
\end{equation}
where $v_{\rm bm}(r|a)$ is the mean relative velocity of all biased tracer-matter pairs separated by $r$ when the expansion factor is $a$ (e.g. Peebles 1980).

On the large scales where linear theory applies,
\begin{equation}
  \xi_{\rm bm}(r|a) 
                  = D_a\,(D_a b_v + b_{\rm L})\,\xi_{\rm mm}(r|a=1).
  \label{xi_largeScale}
\end{equation}
\citep{sdhs01}.  (Strictly speaking, this expression ignores the $k$-dependence of $b_v$ and $b_{\rm L}$; to include it correctly, we must think of $\xi_{\rm mm}$ as an integral over $P(k)$, and include both $b_v$ and $b_{\rm L}$ in the integrand.)  This, in the continuity equation, implies 
\begin{equation}
 [b_v + b_{\rm E}(a)]\,f_a\,\bar\xi_{\rm mm}(r|a) = -\frac{v_{\rm bm}(r|a)}{Hr}\,3\,[1+\xi_{\rm bm}(r|a)],
 \label{v12lin}
\end{equation}
where $f_a\equiv \partial\ln D_a/\partial\ln a$, so that 
\begin{equation}
  v_{\rm bm}(r|a) = 
  \frac{[b_v + b_{\rm E}(a)]}{2}\,v_{\rm mm}(r|a)\,\frac{1+\xi_{\rm mm}(r|a)}{1+\xi_{\rm bm}(r|a)}
\end{equation}
\citep{sdhs01}. This relates the pairwise velocity between biased tracers and the mass to that of the dark matter.  In the current context, it is interesting to write equation~(\ref{v12lin}) as
\begin{equation}
  \frac{v_{\rm bm}(r|a)}{f_a\,Hr}\,[1+\xi_{\rm bm}(r|a)] =
  - \frac{\bar\xi_{\rm bm}(r|a)}{3} - b_v\,\frac{\bar\xi_{\rm mm}(r|a)}{3}.
 \label{v12shds}
\end{equation}
Previous work \citep{HSW_2014,dchp16} has missed the (pair-weighting) term in square brackets on the left hand side, and the second term on the right hand side (the one proportional to $b_v$).  

It is interesting to compare this with the statistical definition \citep[e.g.][]{kf95}:  
\begin{equation}
 v_{\rm bm}(r|a) \equiv \frac{\langle (1+\delta_m)(1+\delta_b)(v_m-v_b)\rangle}{\langle(1+\delta_m)(1+\delta_b)\rangle}.
\end{equation}
In linear theory, velocities and densities at the same position are uncorrelated, so
\begin{align}
  v_{\rm bm}(r|a)
   &\equiv \frac{\langle \delta_m v_b - \delta_bv_m\rangle}{\langle(1+\delta_m)(1+\delta_b)\rangle}
   = \frac{\langle \delta_m b_v v_m - b_{\rm E}\delta_mv_m\rangle}{\langle(1+\delta_m)(1+\delta_b)\rangle}\nonumber\\
   &= \frac{b_v + b_{\rm E}}{2}\, \frac{2\langle\delta_m v_m\rangle}{1+\xi_{\rm bm}}    = f_aHr\,\frac{b_v + b_{\rm E}}{2}\, \frac{2\bar\xi_{\rm mm}/3}{1+\xi_{\rm bm}}\nonumber\\
   &= f_aHr\,\frac{b_v\bar\xi_{\rm mm} + \bar\xi_{\rm bm}}{3\,(1+\xi_{\rm bm})},
\end{align}
where $b_v$ accounts for the fact that void speeds may differ from those of the dark matter.  Note that this agrees with equation~(\ref{v12shds}).  However, this formulation makes it easy to see the effect of assuming voids do not move:  simply set $b_v\to 0$.  Doing so makes $v_{\rm bm}\to \langle\delta_b v_m\rangle = -(Hr)\,f_a\,\bar\xi_{\rm bm}/3$.  However, since voids do move (Figure~\ref{fig:displacements}), there are corrections to this simple expression which should not be ignored if $|b_{\rm E}|$ is not large compared to unity.

Of course, we can go beyond linear theory by inserting equation~(\ref{dEdLcombined}) for $1 + \bar\xi_{\rm bm}$ in equation~(\ref{continuity}). This yields 
\begin{align}
 \label{v12full}
  -\frac{v_{\rm bm}}{f_a\,Hr}\,3\,[1+\xi_{\rm bm}] &= 
  D_a\delta_{\rm L} \,\frac{1+\delta_{\rm Sph}}{1 - D_a\delta_{\rm L}/\delta_c}
  + 2D_a \delta_2 \\
  &= \bar\xi_{\rm bm} + D_a\delta_2 - \delta_{\rm Sph} \nonumber\\
  & \qquad + \delta_c\,(1+\delta_{\rm Sph})\left[(1+\delta_{\rm Sph})^{1/\delta_c} - 1\right] \nonumber
\end{align}
where $v_{\rm bm}$ and $\xi_{\rm bm}$ are on scale $r$, $\delta_{\rm Sph}$ is within $r$, and $\delta_{\rm L}$ is within $R_{\rm L}$.  It is easy to check that, on large scales, this reduces to the linear theory analysis above.  The second equality writes $v_{\rm bm}$ as $\bar\xi_{\rm bm} + D_a\delta_2\ + $ additional terms which can be expressed in terms of $\delta_{\rm Sph} = \bar\xi_{\rm bm} - D_a\delta_2$ (c.f. equation~\ref{dEdLcombined}).  This highlights the terms which are missing from previous analyses.  


\begin{figure}
  \centering
 \includegraphics[width=0.95\hsize]{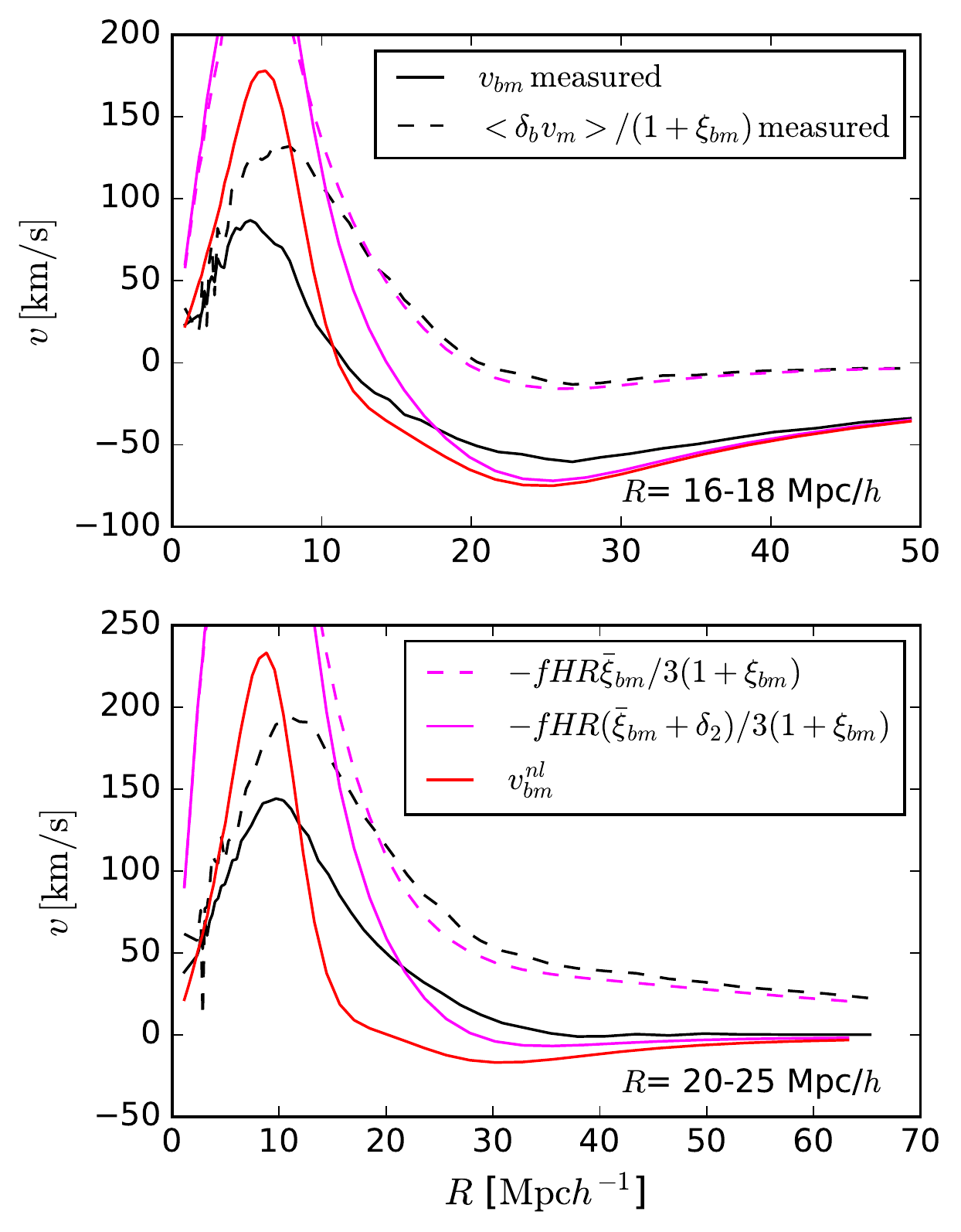}
 \caption{Measured (black) and predicted (magenta and red) pairwise velocity for void-dark matter pairs, as a function of separation at $z=0$, for the same void samples as shown in the bottom two panels of the previous figure.  Dashed curves ignore the contribution from the motion of void centers; solid curves include it.  Magenta curves show the linear theory prediction (equation~\ref{v12shds} with $b_v=0$ or with $b_v$ given by equation~\ref{bvel}), and red curve shows the full signal (equation~\ref{v12full}).  }
 \label{fig:v12}
\end{figure}

\begin{figure}
  \centering
 \includegraphics[width=0.95\hsize]{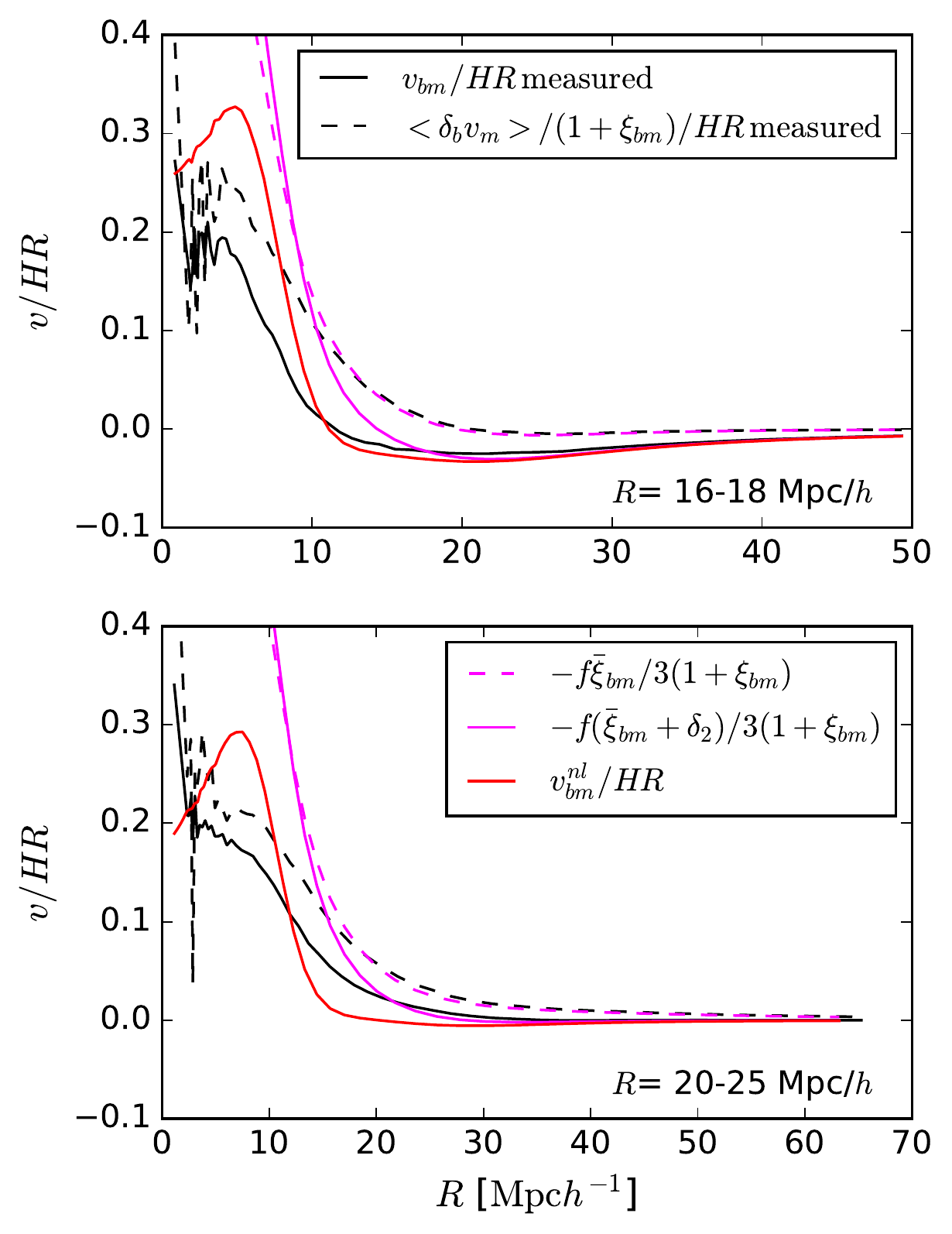}
  \caption{Same as previous figure, but now for $v_{\rm bm}/Hr$.}    
 \label{fig:v12Hr}
\end{figure}

Figure~\ref{fig:v12} compares the measured $v_{\rm bm}$ with the linear and nonlinear predictions, and Figure~\ref{fig:v12Hr} shows a similar comparison of measured and predicted $v_{\rm bm}/Hr$, since this highlights the similarity to $\bar\xi_{\rm bm}$ on large scales.  To illustrate the effect of void motions on the outflow velocity `profile', the solid black curves in Figure~\ref{fig:v12} show 
\begin{displaymath}
  \avg{v_{12}(r)} \equiv
  \frac{\sum_{i=1}^{N_v} \sum_{m=1}^{N_p}{\cal I}_{mi}\,(\mathbf{v}_m - B\mathbf{v}_i)\cdot \mathbf{r}_{mi}/|\mathbf{r}_{mi}|}{\sum_{i=1}^{N_v} \sum_{m=1}^{N_p}\,{\cal I}_{mi}}
\end{displaymath}
with $B=1$, where $N_v$ and $N_p$ are the number of voids and dark matter particles in the simulation box, $\mathbf{r}_{mi}\equiv \mathbf{r}_m-\mathbf{r}_i$ is the separation between void $i$ and particle $m$, and ${\cal I}_{mi}=1$ if $|\mathbf{r}_{mi}|$ is within $\Delta r$ of $r$.  The dashed black curves show this same sum but with $B=0$; this clearly ignores void motions, so the dashed black curves are like the measurements shown in \cite{HSW_2014} (although they used an additional Voronoi weighting scheme).  The dashed and solid curves are different on all scales, showing that void motions matter.

\begin{figure}
  \centering
 \includegraphics[width=0.95\hsize]{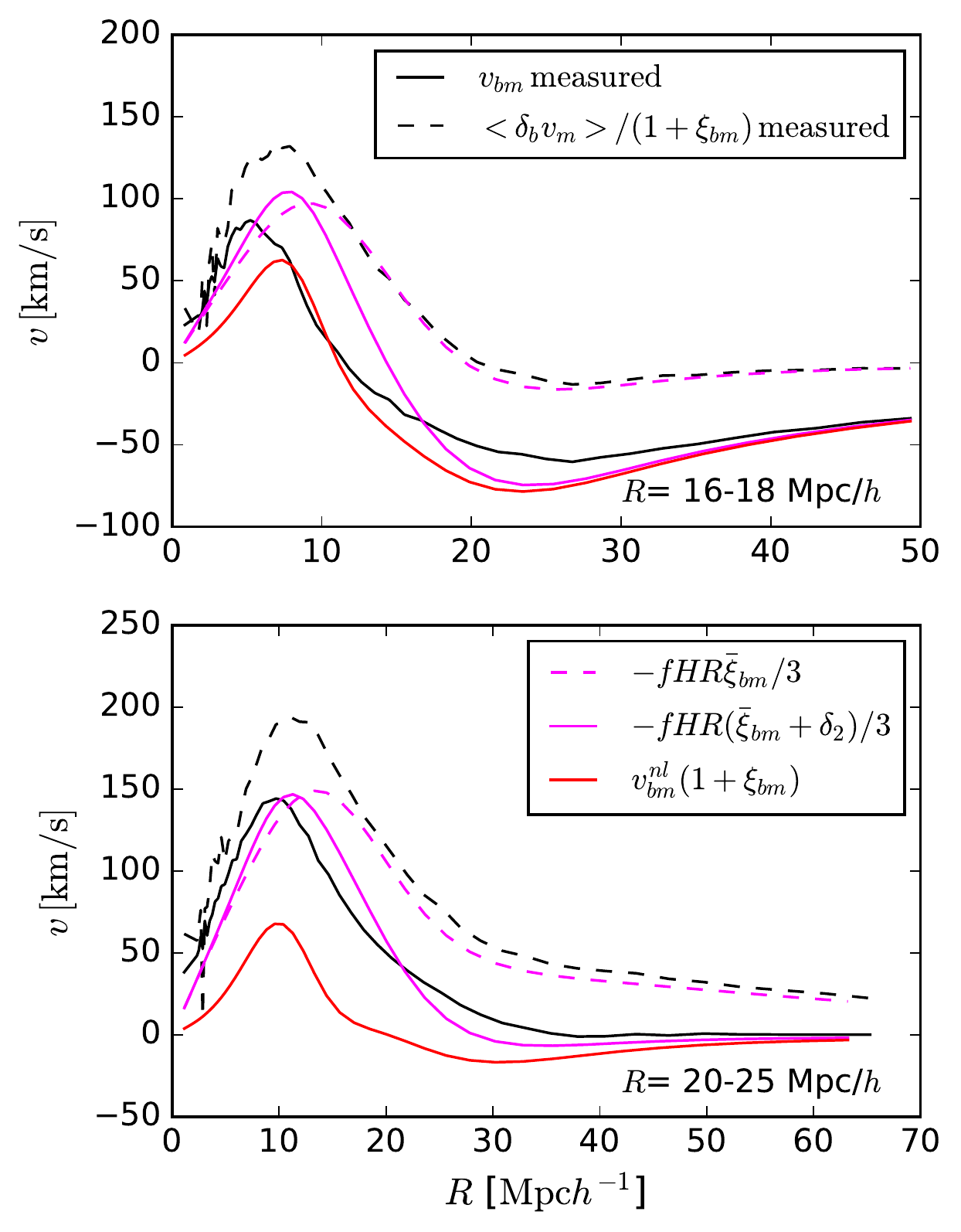}
 \caption{Same as Figure \ref{fig:v12}, but now the predictions have not been divided by $1+\xi_{\rm bm}$.  }
 \label{fig:v12_wrong}
\end{figure}

The corresponding magenta curves show the linear theory predictions:  dashed shows equation~(\ref{v12shds}) with $b_v=0$, and solid uses $b_v\ne 0$ of equation~(\ref{bvel}).  The solid red curve shows the full nonlinear prediction (equation~\ref{v12full}).  While it does not reproduce the measurements accurately on small scales, it does fare better than linear theory.  The agreement with the measurements improves slightly if we do not divide the predictions by $1+\xi_{\rm bm}$ (see Figures~\ref{fig:v12_wrong} and~\ref{fig:v12Hr_wrong}).  This factor, which normalizes the predictions by the number of pairs, as is done for the measurement, is often ignored.  Ignoring it happens to work well for velocities onto clusters as well \citep{sz09}.  

\subsection{Comparison with some previous work}

Before we move on to modeling these results, it is interesting to contrast our treatment of void density and velocity profiles with previous work.

As we noted above, setting $b_v\to 0$ in our equation~(\ref{v12shds}) yields the expression used by \cite{HSW_2014} (though they ignore the $1+\xi_{\rm bm}$ factor).  They report good agreement with their analysis, so it necessary to address why.  Equation~(3) of \cite{HSW_2014} shows that they only sum over particle velocities -- they explicitly do not include the void velocity when measuring the `pairwise' velocity in their simulations.  This means they have unwittingly set $b_v=0$ in their measurements, which is why they find no evidence for this term.  The agreement between the dashed curves in Figure~\ref{fig:v12} shows that, at least on large scales, if the measurement ignores void motions, then one does not do too badly if one ignores them in the theory also.

\cite{dchp16} describe a study of the spherical evolution of cosmic voids; while they find encouraging agreement for voids that are larger than those we have studied, they report that their methodology fails for smaller voids.  Our analysis shows why.  Demchenko et al. use the void center defined at $z=0$ and ask if the spherical model can describe how the profile around this position evolved.  However, voids move (see our Figure~\ref{fig:displacements}), and there is no reason why the spherical model should apply around a position which is not comoving with the center.  E.g., in our Figure~\ref{fig:onlySC}, spherical evolution alone is obviously a bad description of the measured evolution, most especially on large scales, because the measured bias evolves.  Why, then, did they find good agreement with their measurements?  There are two parts to the answer.  

First, their measurements of the void profile at all other $z$ were always centered on the $z=0$ position.  Therefore, their measurements are the real-space analogue of our Figure~\ref{b0}, for which the large scale bias does not evolve!  As a result, the failure of the spherical model on large scales, which is so obvious in our analysis, was hidden by the measurement they chose to model.  
Second, our Figure~\ref{fig:displacements} shows that void motions are approximately independent of void size, so ignoring them is most problematic for small voids.  \cite{dchp16} considered voids with effective sizes that were much larger than the typical void displacements:  presumably this is why they found that agreement with the model degraded as void size decreased.

\section{The Excursion Set Troughs model}\label{sec:ESTmodel}
Having shown that we have a reasonably accurate model for the evolution of the void-matter cross-correlation function, we now ask if its shape is understood.  For this, we use the excursion set troughs model (Section~4.6 in Sheth \& van de Weygaert 2004), with the technical improvements described in Paranjape et al. (2013).  Briefly, although the approach is most often used to predict abundances of objects from the statistics of the initial fluctuation field, it actually includes a model for the Lagrangian space profiles around the protohalo or protovoid patches. This is a direct consequence of recognizing the often overlooked fact that the cross-correlation between the patches and the mass {\em is} the mean density profile \citep{bbks86,rks98}, and, for the excursion set approach, this Lagrangian cross-correlation is known \citep{ms12,mps12}.

In what follows, we first write down the Lagrangian cross-correlation function, and show that it provides a reasonable fit to the Lagrangian measurements.  We then insert the Lagrangian profile in equation~(\ref{dEdLcombined}) and fit it to the evolved Eulerian profile.  Agreement between the fitted coefficients in Lagrangian and Eulerian space is an indication that all is self-consistent.  

\begin{figure}
  \centering
 \includegraphics[width=0.95\hsize]{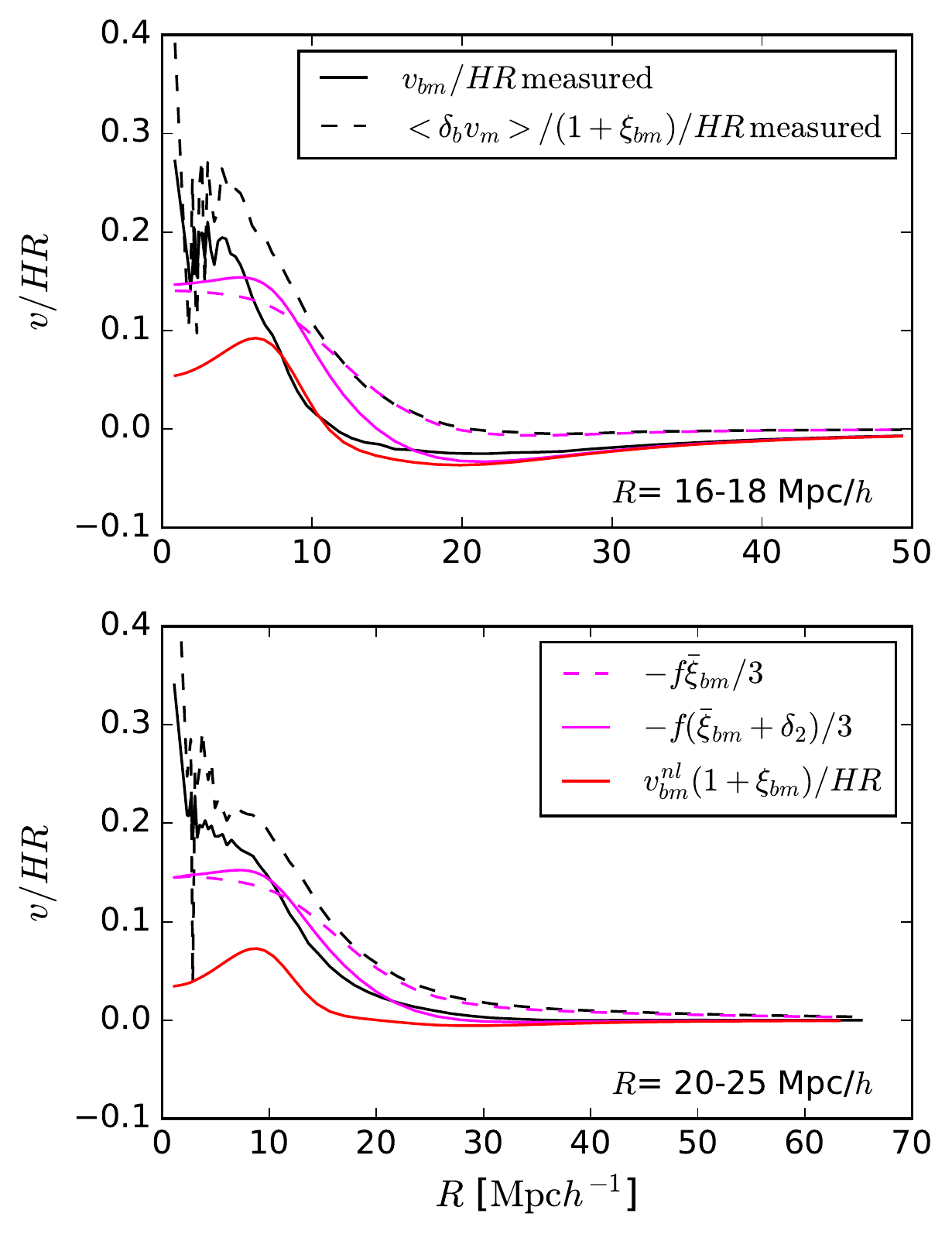}
  \caption{Same as Figure~\ref{fig:v12Hr}, but now the predictions have not been divided by $1+\xi_{\rm bm}$.}    
 \label{fig:v12Hr_wrong}
\end{figure}

\subsection{EST Lagrangian protovoid profiles}
In Appendix~A we discuss why the Lagrangian cross-correlation function -- the density profile centered on positions which are Excursion Set Troughs on scale $R_p$ (EST$_p$) -- is well-approximated by 
\be
 \delta_{\rm L}(<R_q|{\rm EST}_p) = b_{10}\,s_0^{pq}
              + b_{01}\,2\frac{\dd s_0^{pq}}{\dd\ln s_0^{pp}},
 \label{dprofile}
\ee
where $s_0^{pp}$ and $s_0^{pq}$ were defined earlier (equation~\ref{sjpq}), and 
\be
 b_{10} + b_{01} = \frac{\avg{\!\delta_p\!}}{s_0^{pp}}
 \label{bconsistency}
 \ee
 where $\avg{\!\delta_p\!}$ denotes the typical value of the Lagrangian overdensity enclosed within a protovoid of size $R_p$ (Castorina et al. 2017; Chan et al. 2017).  Usually, this value is known: e.g., the simplest expectation for voids is that it equals $-2.7$.  However, in the present case, we do not have an a priori prediction for $\avg{\!\delta_p\!}$ because, as we noted in our discussion of the spherical model, this value is rather sensitive to the details of the void finder \citep[also see][]{Nadathur:2015}.  Indeed, not only is it not known, but, in contrast to the EST model in which the mean Lagrangian profile is obtained by stacking voids of fixed Lagrangian size, here, we stack voids based on their Eulerian rather than Lagrangian size.  Therefore, in what follows, we will treat $b_{10}$ and $b_{01}$ as free parameters, which will be determined by fitting to the measured Lagrangian profile.  We will then check if the fitted values do indeed estimate $\delta_p$ well (following equation~\ref{bconsistency}).  However, predicting how these parameters depend on void size is beyond the scope of this work.

In practice, we do not fit to the solid curves in Figure~\ref{fig:dLag} because of the problems on (small) scales that are comparable to the interparticle separation.  Rather, we fit to the dot-dashed curves shown there.  These were predicted by our evolution model from the measured profile at $z=2$; they provide an excellent description of the Lagrangian profile, but do not suffer from artifacts on small scales.  In addition, when fitting, we set $R_p=R_{eff}/2$, where $R_{eff}$ is the scale shown in the legend in the upper left of the figure.  This sets the values of $s_0^{pp}$ and $s_0^{pq}$.  

\begin{table}
\begin{center}
\begin{tabular}{| c | c | c | c | c | c |}
\hline
$R_{\rm eff}$ [$h^{-1}$Mpc] & $b_{10}$ & $b_{01}$ & $s_0^{pp}$ & $s_1^{pp}$ & $\delta_p$ \\
\hline
$14-16$ & $-0.54$ & $-0.72$ & 0.71 & 0.03 & $-0.90$ \\
\hline
$16-18$ & $-0.75$ & $-0.95$ & 0.60 & 0.02 & $-1.02$ \\
\hline
$20-25$ & $-1.5$ & $-1.7$ & 0.39 & 0.01 & $-1.26$ \\
\hline
\end{tabular}
\caption{\label{tab:lag_fit}Results from fitting equation~(\ref{dprofile}) to the Lagrangian profile.}
\end{center}
\end{table}

\begin{figure}
 \centering
 \includegraphics[width=0.95\hsize]{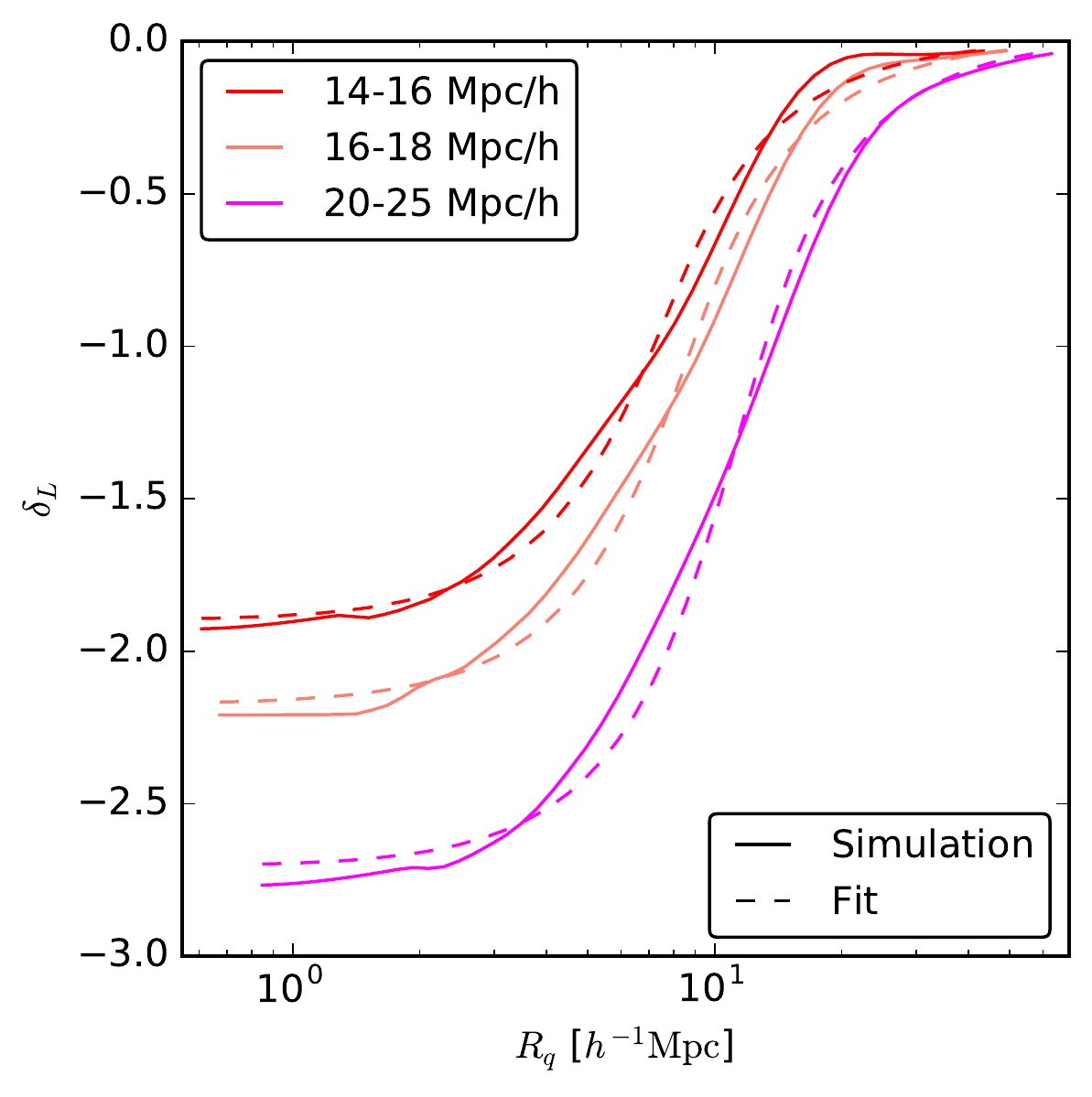}
 \caption{Profiles of the form given by \eqn{dprofile} (dashed) with free parameters $b_{10}$ and $b_{01}$ determined by fitting to the Lagrangian profile (solid, same as dot-dashed in Figure~\ref{fig:dLag}).}
 \label{fig:L_profile_fit}
\end{figure}

Table~\ref{tab:lag_fit} gives the best-fitting parameters; the fits themselves are shown as dashed lines in Figure~\ref{fig:L_profile_fit}.  Although the agreement with the measured solid curves is not spectacular -- the predicted shape is flatter than measured on the inner side of the void wall -- it does suggest that the excursion set troughs approach provides a reasonably good framework within which to describe voids.  As a final consistency check, the final column of Table~\ref{tab:lag_fit} gives the value of $\delta_p$ predicted by equation~(\ref{bconsistency}).  It is in rather good agreement with $\delta_{\rm L}(<R_p)$ on the scale $R_p=R_{\rm eff}/2$ shown in the Figure.  

\subsection{EST + spherical evolution + extra term}
We now test if our ESP model for the Lagrangian profile shape, when combined with equation~(\ref{dEdLcombined}) for the evolution, is able to provide a good description of the Eulerian profiles in the simulation.  As before, we keep the structure of the model fixed, but fit for the coefficients $b_{10}$ and $b_{01}$.  If these turn out to be the same as those in Table~\ref{tab:lag_fit}, then this will be an indication that our approach is self-consistent.

\begin{figure}
 \centering
 \includegraphics[width=0.95\hsize]{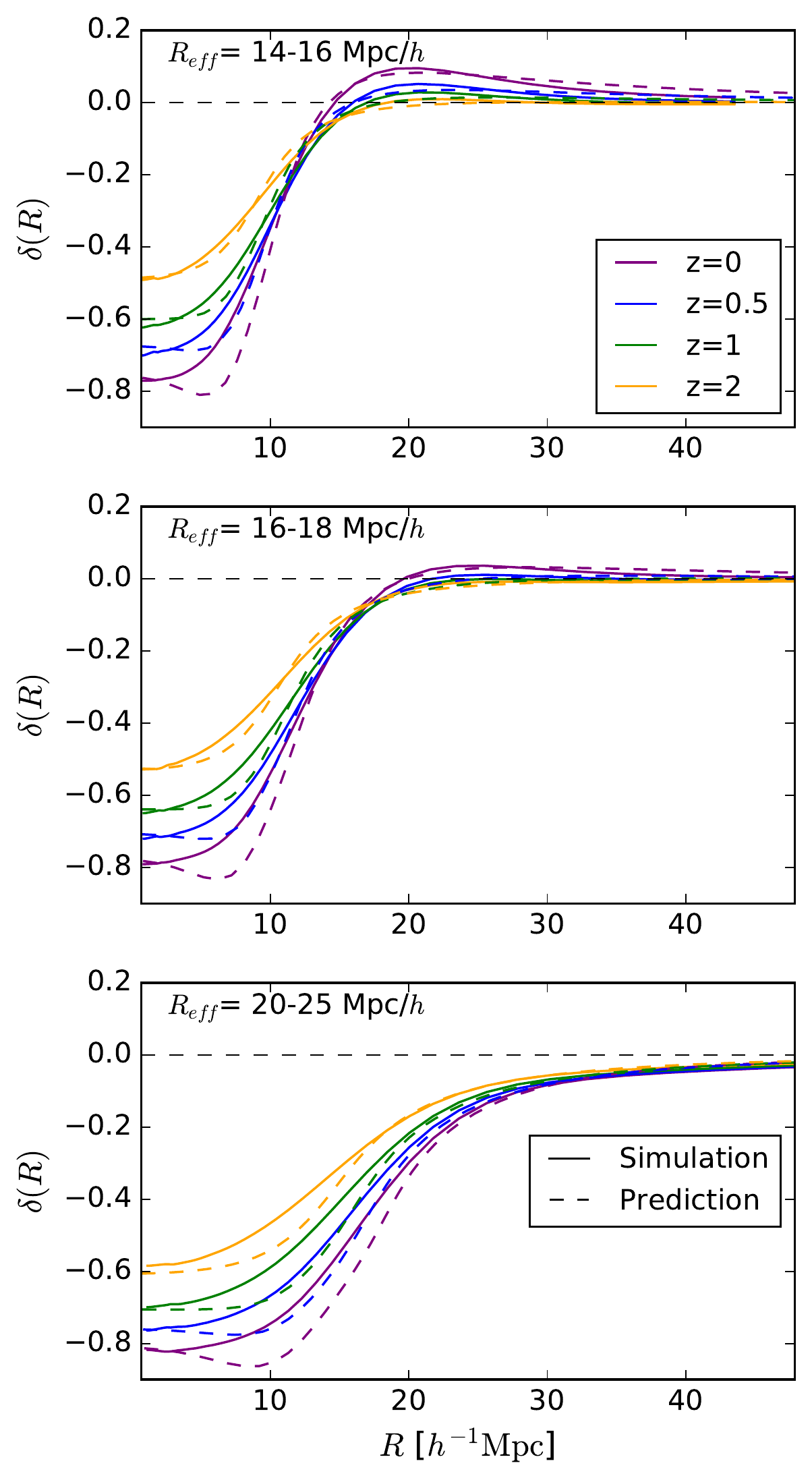}
 \caption{Profiles of the form given by inserting \eqn{dprofile} in \eqn{dEdLcombined} (dashed) with free parameters $b_{10}$ and $b_{01}$ determined by fitting to the Eulerian profile (solid, same as Figure~\ref{fig:complete}).}
 \label{fig:E_profile_fit}
\end{figure}

\begin{figure}
 \centering
 \includegraphics[width=0.95\hsize]{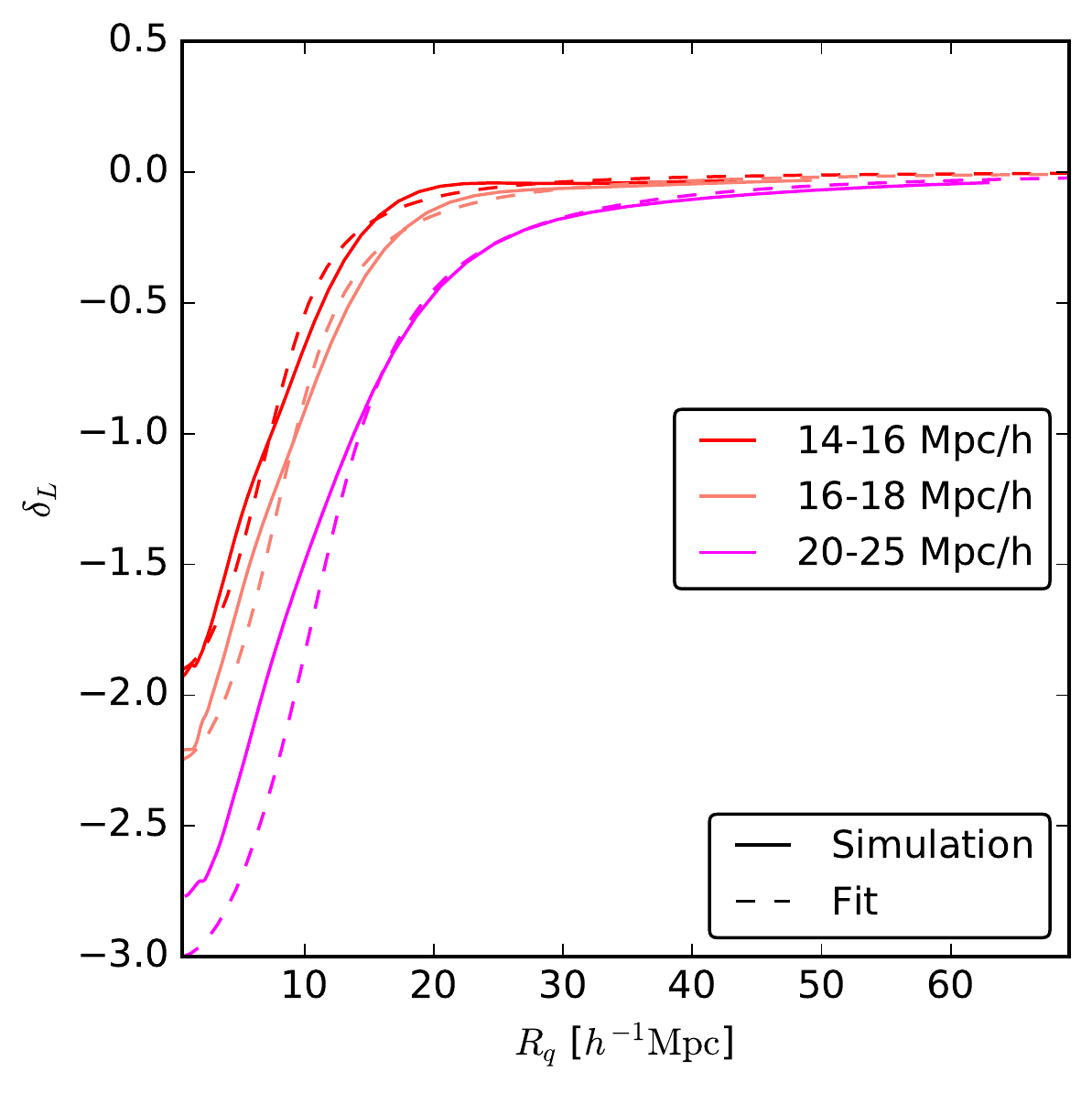}
 \caption{Profiles of the form given by \eqn{dprofile} (dashed) with free parameters $b_{10}$ and $b_{01}$ determined by fitting to the Eulerian profile (solid, same as dot-dashed in Figure~\ref{fig:dLag}).}
 \label{fig:Efit2L}
\end{figure}

Figure~\ref{fig:E_profile_fit} shows the result of fitting the Eulerian profiles (solid, same as Figure~\ref{fig:complete}) with the shape that is obtained by inserting \eqn{dprofile} in \eqn{dEdLcombined} (dashed) and fitting for $b_{10}$ and $b_{01}$.  In practice, because evolution and averaging do not commute (Figure~\ref{fig:noCommute}) we restrict the fit to scales $R<R_{eff}/5.5$ and $R>R_{eff}$. The best-fit values are listed in Table~\ref{tab:eul_fit}. Comparison with those in Table~\ref{tab:lag_fit} shows reasonable agreement, especially for large voids.

Figure~\ref{fig:Efit2L} shows another test of self-consistency:  the dashed curves show the result of using the $b_{10}$ and $b_{01}$ values obtained from fitting the Eulerian profiles to predict the measured Lagrangian profiles (solid).  The agreement is again reasonable, except around the void wall where we know non-commutation matters, and where the predicted profiles tend to be slightly shallower than the measured ones.  We conclude that equation~(\ref{dEdLcombined}) provides a reasonably flexible and physically motivated `universal' framework for fitting void profiles.

\begin{table}
\begin{center}
\begin{tabular}{| c | c | c | c | c | c |}
\hline
$R_{\rm eff}$ [$h^{-1}$Mpc] & $b_{10}$ & $b_{01}$ & $s_0^{pp}$ & $s_1^{pp}$ & $\delta_p$ \\
\hline
$14-16$ & $-0.29$ & $-0.98$ & 0.63 & 0.03 & $-0.80$ \\
\hline
$16-18$ & $-0.50$ & $-1.31$ & 0.60 & 0.02 & $-0.96$ \\
\hline
$20-25$ & $-1.3$ & $-2.4$ & 0.34 & 0.01 & $-1.27$ \\
\hline
\end{tabular}
\caption{\label{tab:eul_fit}Results from fitting~\eqn{dEdLcombined} with $\delta_L(<R_L)$ from ~\eqn{dprofile} to the Eulerian profile.}
\end{center}
\end{table}

\subsection{Scale-dependence of bias}\label{sec:bk}
Before we conclude this section, it is interesting to contrast the role played by the extra term, which, on large scales, makes
\be
 b_{10}\to  b_{10} + D_a \qquad{\rm and}\quad
 b_{01}\overset{\!\!\sim}{\to}  b_{01} - D_a,
 \label{b10Eulb01Eul}
\ee
so that $b_{10} + b_{01}$ of the evolved profile is approximately independent of time, and equation~(\ref{bconsistency}) is a reasonable approximation at all times.  (The Appendix discusses why the expression for $b_{01}$ is only an approximation.)

For massive halos, $b_{10}$ and $b_{01}$ are both positive, and increase with halo mass.  So the term from evolution increases the large scale bias and decreases the amplitude of the scale-dependence term:  as a result, bias appears to be scale-independent over a wider range of scales in Eulerian than Lagrangian space.  For lower mass halos, $b_{10}$ can be negative, but $b_{01}$ will still be positive (because of equation~\ref{bconsistency}).  In this case, evolution will bring both closer to zero, and may even make both change sign.  Thus, generically, evolution makes Eulerian bias less scale dependent than Lagrangian.  

In contrast, for large voids, $b_{10}$ and $b_{01}$ are both negative.  Evolution makes the scale-independent piece closer to zero, but makes the scale-dependent piece even more negative, with the result that Eulerian void bias is more scale-dependent than Lagrangian.  This is also true for small voids:  although $b_{10}$ can be positive (but $b_{01}$ negative), evolution will increase the scale-independent piece, and make $b_{01}$ more negative.  Again, Eulerian void bias is more scale dependent than Lagrangian -- unlike for halos.

\section{Discussion}\label{sec:discuss}
We studied void density and velocity profiles -- the former being just the void-matter cross correlation function -- in simulations.  We showed that the cross-correlation function evolves both because of how matter flows around (away from) the voids (equation~\ref{dEdL}), but also because voids move.  Voids identified today (and by extension, voids identified at any given time) are displaced from their initial protovoid positions by an amount that is approximately independent of void size (Figure~\ref{fig:displacements}).  These displacements make the void-matter cross-correlation evolve even on very large scales (Figures~\ref{b0} and~\ref{b99}), in agreement with theory (equation~\ref{bEbL}).  Accounting for these displacements (equation~\ref{dEdLcombined}) is necessary to explain the evolution of void density profiles (compare Figure~\ref{fig:onlySC} with~\ref{fig:complete}) as well as the profiles of velocity inflow/outflow around voids (Figures~\ref{fig:v12}--\ref{fig:v12Hr_wrong}).  These displacements and their consequences are missing from all previous work on void evolution.  The implications for redshift space distortions are the subject of work in progress.

We then explored if the simplest Excursion Set Troughs model provides a useful `universal' framework for describing void profiles.  This framework ignores details associated with non-spherical evolution \citep{smt01,scs13,anp15}, and the fact that evolution and averaging do not commute (Figure~\ref{fig:noCommute}), although the non-commutation matters most on scales that are within the void boundary.  Although the EST framework is, in principle, fully predictive, predicted void abundances are rather sensitive to how voids are characterized in the evolved field (see discussion associated with Figure~\ref{fig:SC}).  To study if the EST framework is useful even when the details of the void finder are not understood, we treated the EST expression for void profiles (equation~\ref{dprofile}) as a framework with free parameters which are to be determined by fitting to data.  This worked reasonably well.  If one evolves the measured void profile back to the initial Lagrangian conditions and fits there (Figure~\ref{fig:L_profile_fit}), then the fitted parameters yield reliable information about the protovoid patches from which the voids evolved (equation~\ref{bconsistency}).  Moreover, these Lagrangian parameters can also be determined by inserting equation~(\ref{dprofile}) in equation~(\ref{dEdLcombined}) and fitting to the evolved Eulerian profile (Figures~\ref{fig:E_profile_fit} and~\ref{fig:Efit2L}).  Fitting to either the Lagrangian or the Eulerian profiles yields similar estimates of the Lagrangian bias parameters $b_{10}$ and $b_{01}$, which describe the scale-dependence of void bias (compare Tables~\ref{tab:lag_fit} and~\ref{tab:eul_fit}).

When combined with equation~(\ref{bconsistency}), these values indicate that protovoid patches which grow into voids identified by VIDE have average enclosed overdensities of order $-1$, rather than the value of $-2.7$ that is predicted by theory.  Much of this difference is because the void walls are not sharp, so the thickness of the wall influences its estimated size.  Nevertheless, it is interesting that a Lagrangian underdensity of order $-1$ is close to the value suggested by fitting void abundances using the EST approach.  Exploring if the EST approach allows self-consistent determinations of void profiles and abundances is left to future work.

Future work must also address the question of how to model voids identified using biased tracers, rather than dark matter particles, in the evolved field (as we have done here).  If $\delta_g$ denotes the overdensity of the bias tracer, then the simplest approximation would simply set $1+\delta_{\rm E} = 1 + \delta_g/b_g$ in all the expressions of this paper.  Section~\ref{sec:bk} motivates why scale-independent linear bias may be a better approximation for the halos which host galaxies in the evolved field than it is for voids, so that the simple prescription above may work reasonably well in practice.  




\section*{Acknowledgments}


We are grateful to the participants of the meeting on voids at the Simons CCA at the end of September 2018 for discussions which encouraged us to publish our findings.  Our work on this project began in 2013.  RKS is grateful to the Perimeter Institute for its hospitality in September 2013, the Institut Henri Poincare for its hospitality in November 2013, and the members of the HECAP group at ICTP for their understanding during this time, as well as their hospitality since.  He is also grateful to the IPMU for hospitality in March 2014, when he presented many of these results.  This work has made use of the python pylians libraries, publicly available at https://github.com/franciscovillaescusa/Pylians. EM is grateful to Francisco Villaescusa-Navarro for creating these libraries. She acknowledges the WFIRST Science Investigation Team “Cosmology with the High Latitude Survey”  supported by NASA grant 15-WFIRST15-0008.  

\bibliography{mybib}{}

\appendix

\section{Bias in the excursion set troughs approach}\label{ESTbias}
This Appendix is intended to provide some insight into the structure of \eqn{dprofile}.  Here, we will use the notation $\avg{\!\bar\delta_q|{\rm EST}_p\!}$ to denote the average value of the overdensity within $R_q$, given that, when smoothed on scale $R_p$, there is an Excursion Set Trough at the center.  I.e., this is the quantity that was denoted $\delta_{\rm L}(<R_q)$ in the main text.

\subsection{Which variables matter}
In general, there are as many bias factors as there are variables which matter for defining what makes the protovoid.  For example, if the only condition for making a void is that $\bar\delta_p$ must equal a certain value (say $-2.7$), then $b_{10}= -2.7/s_0^{pp}$ and $b_{01}=0$.  Here $b_{01}$ is zero because only one variable (in this case $\bar\delta_p$) matters.  If the condition is that $\bar\delta_p$ must exceed a threshold, $\bar\delta_p\le -2.7$ say, then $b_{01}$ is again zero (since, again, only one variable matters), but $b_{10}$ is now a more complicated function of $-2.7/\sqrt{s_0^{pp}}$ (see equation~ in Castorina et al. 2017), since now $\avg{\!\bar\delta_p\!}\ne -2.7$.

In the upcrossing approximation to the excursion set approach \citep{ms12}, $\bar\delta_p$ must exceed a threshold on scale $R_p$, but must lie below it on the next larger scale:  $\bar\delta_p \le -2.7$ and $\bar\delta_{p+\Delta} \ge -2.7$.  In effect, the second condition constrains ${\rm d}\bar\delta_p/{\rm d}R_p \ge 0$.  Since ${\rm d}\bar\delta_p/{\rm d}R_p$ is a different variable from $\bar\delta_p$ itself, in this case there are two bias factors:  both $b_{10}$ and $b_{01}$ may be nonzero.  Since they multiply factors which depend differently on scale, the bias is said to be `scale-dependent' \citep{mps12}.  

The peaks model of \cite{bbks86} does not constrain ${\rm d}\bar\delta_p/{\rm d}R_p$, but it does constrain derivatives of $\bar\delta_p$ with respect to spatial position, so profiles in this model can also be written in terms of two bias factors.  In the excursion set peaks/troughs approach, there are constraints on the spatial derivatives in addition to the excursion set conditions on $\bar\delta_p$ and ${\rm d}\bar\delta_p/{\rm d}R_p$ \citep{ps12,psd13}.  In principle, therefore, this approach comes with three bias factors:  \cite{css17} show why equation~\ref{dprofile} remains an excellent approximation.  Of course, the coefficients $b_{10}$ and $b_{01}$ for EST are different functions of $-2.7/\sqrt{s_0^{pp}}$ than they are in the BBKS model or in the upcrossing approximation.  This means that the associated profile shapes differ for the three approximations.

\begin{figure}
 \centering
 \includegraphics[width=0.9\hsize]{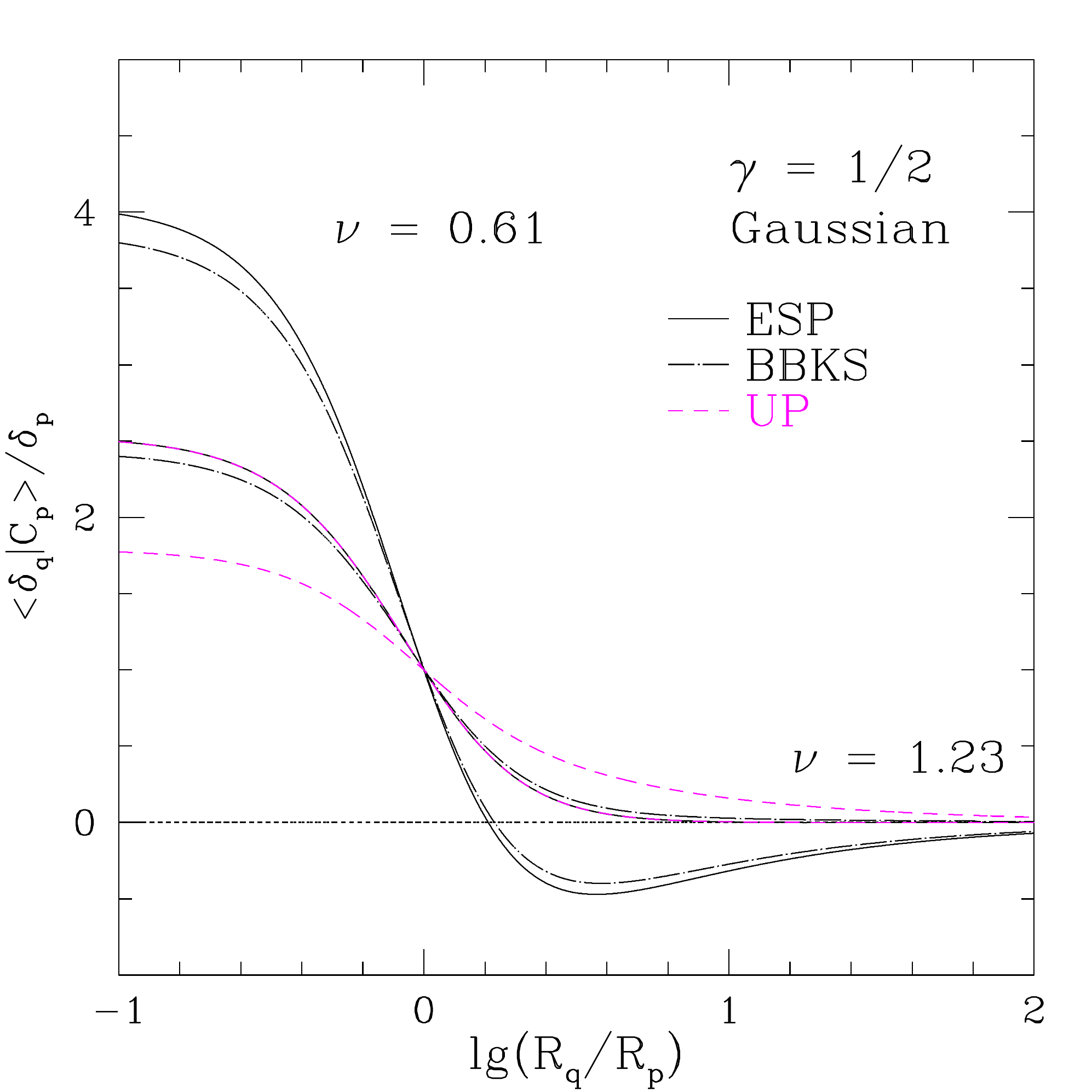}
 \caption{Mean enclosed density profiles (solid) around excursion set peaks in a Gaussian smoothed field having $P(k)\propto k^{-7/3}$ for a range of choices of peak height $\nu$.  Dot-dashed curves show the peak profile of \cite{bbks86} which ignores the upcrossing constraint, and short dashed curves come from requiring upcrossing, but dropping the peaks constraint.  In all cases, lower peaks have steeper profiles, with the trend being most pronounced for ESP. }
 \label{denESP}
\end{figure}

Figure~\ref{denESP} illustrates that, generically, the profiles of high peaks (or extreme troughs) do not cross zero, whereas those of low peaks (or less extreme troughs) do.  In the current context, if all voids had the same Lagrangian underdensity on scale $R_p$, then the large voids would be more extreme, so we would expect them to have shallower profiles than smaller voids.  If nonlinear evolution preserves this correlation, then this would explain the similar trend seen in the evolved profiles of \cite{HSW_2014}.  In practice, some of the protovoids with high walls will not survive the void-in-cloud crushing process; since these have the most dramatic differences in profile shape, removing them from the list of putative protovoids will serve to reduce the dependence of protovoid shape on void size.  (Doing this correctly requires averaging the evolved profiles, rather than evolving the average.)  The picture is further complicated by the fact that, in the main text, we find that larger voids tend to be slightly more {\em under}dense (Figure~\ref{fig:L_profile_fit}); this weakens the correlation between profile slope and void size.  That said, it is worth bearing in mind that the largest voids in our simulations (effective radii of about 25$h^{-1}$Mpc) are amongst the smaller voids in \cite{HSW_2014}.  For void sizes larger than this, \cite{HSW_2014} see qualitatively the same scaling as we predict:   larger voids have shallower profiles and are less empty in their central regions.  

Since there is no theory for why the typical void depths are $\sim -1$ rather than $\sim -2.7$, nor for how void depth should scale with void size, and different authors find different trends for this scaling, we have chosen to {\em not} try to predict either the characteristic density or the dependence on void size.  Rather, in the main text we simply assume that the structure of \eqn{dprofile} is accurate, but we treat $b_{10}$ and $b_{01}$ as free parameters.  This corresponds to assuming that we have identified $\bar\delta_p$ and its derivative with respect to scale $R_p$ as being the most important variables which matter for void formation, but we are agnositic as to exactly what value these variables should have.  

\subsection{Other intuitive rearrangements}
It is useful to use \eqn{bconsistency} to rewrite \eqn{dprofile} so that $b_{10}$ does not appear explicitly:
\begin{align}
 \avg{\!\bar\delta_q|{\rm EST}_p\!}
   & = \delta_p\,\frac{s_0^{pq}}{s_0^{pp}}\, - b_{01}\, s_0^{pq} 
     \left(1 - 2\frac{\dd\ln s_0^{pq}}{\dd\ln s_0^{pp}}\right).
 \label{dscaled}
\end{align}
The first term on the right hand side has the same shape as the density run around random positions of height $\delta_p$; the second term is the correction which comes from the additional excursion set constraint.

Instead, writing it so that $b_{01}$ does not appear explicitly yields
\be
 \avg{\!\bar\delta_q|{\rm EST}_p\!} = \delta_p\,2\frac{\dd s_0^{pq}}{\dd s_0^{pp}}
  + b_{10}\,s_0^{pq}\left(1 - 2\frac{\dd\ln s_0^{pq}}{\dd\ln s_0^{pp}}\right).
 \label{dp1h2h}
\ee
We argue shortly that the second term is the usual linear bias term which dominates on large scales, whereas the first should be thought of as a shot-noise like term which is negligible when $R_q\gg R_p$.  

In both expressions, the quantity $1 - 2\,\dd\ln s_0^{pq}/\dd\ln s_0^{pp}$ plays a key role.  Below, we provide analytic expressions for it for some power spectra and filters; these show that it tends to unity when $R_q\gg R_p$, so both \eqn{dp1h2h} and \eqn{dscaled} indicate that $\avg{\!\bar\delta_q|{\rm EST}_p\!} \to b_{10}(\nu)\,s_0^{pq}$ in this limit.

Finally, the main text argued that, on large scales, the evolved Eulerian profile should be given by adding $D_a\delta_2$ of equation~(\ref{d2}) to this Lagrangian profile.

\subsection{Useful analytic expressions for Gaussian and TopHat filters}\label{analytics}

For Gaussian filtering,
\begin{equation}
 \avg{\!\bar\delta_q|{\rm EST}_p\!} = b_{10}\,s_0^{pq} + b_{01}\,\frac{s_0^{pp}}{s_1^{pp}}\,s_1^{pq}
\end{equation}
for all $P(k)$.  Moreover, for Gaussian filtering, $\delta_2$ of equation~(\ref{d2}) equals $D_a \,(s_0^{pq} - s_1^{pq}s_0^{pp}/s_1^{pp})$,
so 
\be
 \avg{\!\bar\delta_{\rm E}|{\rm EST}_p\!} = (b_{10}+D_a)\,s_0^{pq} + (b_{01}-D_a)\,\frac{s_0^{pp}}{s_1^{pp}}\,s_1^{pq}.
 \ee
I.e., in this case, equation~(\ref{b10Eulb01Eul}) of the main text is exact.

For power-law $P(k)\propto k^n$, 
\be
 \frac{s_0^{pq}}{s_0^{pp}} = \left(\frac{2}{R_q^2/R_p^2 + 1}\right)^{(3+n)/2}
\ee
and
\be
 1 - 2\,\frac{\dd\ln s_0^{pq}}{\dd\ln s_0^{pp}} = 
 \frac{(R_q/R_p)^2 - 1}{(R_q/R_p)^2 + 1},
\ee
so 
\be
 \frac{\avg{\!\bar\delta_q|{\rm EST}_p\!}}{\delta_p} =
  \left(\frac{2}{R_q^2/R_p^2 + 1}\right)^{\frac{5+n}{2}}
  + \frac{b_{10}}{\delta_p/s_0^{pq}}\,
    \left(\frac{1 - (R_p/R_q)^2}{1 + (R_p/R_q)^2}\right).
\ee
This shows that when $R_q\gg R_p$ then the `1-halo' term, the first term on the right hand side, falls as $(R_q/R_p)^{-5-n}$; the second term only falls as $(R_q/R_p)^{-3-n}$, so it dominates on large scales.  We used these expressions to produce Figure~\ref{denESP}.

For a TopHat, these quantities depend differently on $n$.  When $n=-2$, as for Figure~\ref{fig:dvoidTH} in the main text, then 
\be
 \frac{s_0^{pq}}{s_0^{pp}} = 
  \begin{cases} 
    \frac{5 - (R_p/R_q)^2}{4(R_q/R_p)} & \mbox{if } R_q/R_p\ge 1 \\
    \frac{5 - (R_q/R_p)^2}{4}            & \mbox{if } R_q/R_p < 1 
  \end{cases},
 \label{SxTHn-2}
\ee
making 
\be
 1 - 2\frac{\dd\ln s_0^{pq}}{\dd\ln s_0^{pp}} = 
  \begin{cases} 
    \frac{(R_q/R_p)^2 - 1}{(R_q/R_p)^2 - 1/5} & \mbox{if } R_q/R_p\ge 1 \\
    \frac{(R_q/R_p)^2 - 1}{1 - (R_q/R_p)^2/5} & \mbox{if } R_q/R_p < 1 
  \end{cases},
\ee
and so 
\be
 \frac{s_0^{pq}}{s_0^{pp}}\left(1 - 2\frac{\dd\ln s_0^{pq}}{\dd\ln s_0^{pp}}\right) = 
  \begin{cases} 
    \frac{1 - (R_p/R_q)^2}{(4/5)(R_q/R_p)} & \mbox{if } R_q/R_p\ge 1 \\
    \frac{(R_q/R_p)^2 - 1}{4/5} & \mbox{if } R_q/R_p < 1 .
  \end{cases}.
\ee
This has some intuitive appeal, since a little algebra shows that
\begin{align}
  \avg{\!\bar\delta_q|{\rm EST}_p\!}
  &= b_{01}s_0^{pp}\left(\frac{R_p}{R_q}\right)^3 + b_{10}\,s_0^{pq}
  \quad \mbox{if } R_q/R_p\ge 1 \nonumber\\
  &= \delta_p\,\left(\frac{R_p}{R_q}\right)^3
    + \frac{5}{4}\,s_0^{qq}\, b_{10}\, \left(1 - \frac{R_p^2}{R_q^2}\right).
 \label{1h2h}
\end{align}
If we think of the left hand side as a cross-correlation function, then the first term on the right hand side is the one-halo term -- the contribution from the fact that the enclosed overdensity within $R_p$ is $\delta_p$ -- so it matters little on large scales where $R_q\gg R_p$.  The factor of $5/4$ is the $R_q\gg R_p$ limit of $s_0^{pq}/s_0^{qq}$, so the second term, which represents the two-halo contribution, gives the scale-dependence of the linear bias factor, and goes to zero when $R_q\to R_p$.  

We can see this even more clearly if we consider the density at $R_q$ rather than within $R_q$.  For TopHat smoothing of $P(k)\propto k^{-2}$, the overdensity in the shell of radius $R_q$ is 
\be
 \frac{\avg{\!\delta_q|{\rm EST}_p\!}}{\delta_p}
    = \frac{5}{2} \left(1-\frac{R_q^2}{R_p^2}\right) 
             + \frac{\delta_p\,b_{10}}{\nu_p^2}
               \frac{5}{4}\left(\frac{5R_q^2}{3R_p^2} - 1\right)
 \label{n-2TH}
\ee
if $R_q$ is smaller than $R_p$, whereas it is 
\be
 \avg{\!\delta_q|{\rm EST}_p\!} = b_{10}\,\xi(R_q)
\ee
when $R_q$ is larger than $R_p$.  (Here $\delta_q$ has no bar, to indicate that it is the density at $R_q$ rather than within $R_q$.)   
Evidently, for this case, there is no scale dependence to bias, except for the sharp cutoff due to exclusion on scales smaller than $R_p$.  I.e., the second term in \eqn{1h2h} is the simplest possible two-halo term:  it is just $(3/R_q^3)\int_{R_p}^{R_q} \dd r\,r^2\,b_{10}\,\xi(r)$, upon noting that $\xi(R_q) = 5s_0^{qq}/6$.

The main text (equation~\ref{dEdLcombined}) argued that evolution means we must add a term which is proportional to $\delta_2$ of equation~(\ref{d2}).  For TopHat smoothing of $P(k)\propto k^{-2}$ this term is 
\be
 \delta_2 = 
  \begin{cases}
   s_0^{pq} - s_0^{pp} (R_p/R_q)^3 & \qquad {\rm if}\quad R_q\ge R_p\\
   s_0^{pq} - s_0^{pp}            & \qquad {\rm if}\quad R_q< R_p
  \end{cases}.
 \ee
Therefore, on scales where $R_q/R_p\ge 1$, the evolved profile is 
\be
  \avg{\!\bar\delta_{\rm E}|{\rm EST}_p\!}
  = (b_{01}-D_a)\,s_0^{pp}\left(\frac{R_p}{R_q}\right)^3 + (b_{10}+D_a)\,s_0^{pq}.
 \label{1h2hEul}
\ee
In this case, as for Gaussian smoothing, equation~(\ref{b10Eulb01Eul}) of the main text is exact.  Therefore, $b_{10}^{\rm E} + b_{01}^{\rm E} = b_{10}+b_{01}$ is invariant, and satisfies equation~(\ref{bconsistency}) at all times.


Unfortunately, this is not generic.  For example, it is a simple matter to write down the appropriate expressions for the Markov Velocity models of \cite{ms14mv}.  These models can be associated with smoothing filters that are compact in Fourier space:  $W_\alpha(kR) = [1 - (kR)^\alpha]\vartheta(1 - kR)$.  For scale free power spectra, these models are characterised by one parameter, $0\le \gamma\le 1$, which satisfies
\be
 1/2\gamma^2 = 1 + \alpha/(3+n).
\ee
These models have 
\be
  \avg{\!\bar\delta_q|{\rm EST}_p\!}
  = b_{10}\,s_0^{pq} + b_{01}s_0^{pp}\left(\frac{s_0^{qq}}{s_0^{pp}}\right)^{1/2\gamma^2}
  \quad \mbox{if } R_q/R_p\ge 1 ,
 \label{mrkvs}
\ee
so their structure is quite similar to TopHat smoothing of $n=-2$.  
However, the integral of $b_v$ yields
\begin{displaymath}
 s_0^{pq} - s_0^{pp}\, \left(\frac{s_0^{qq}}{s_0^{pp}}\right)^{(5+n)/(3+n)}\,
                       \frac{s_1^{pq}}{s_1^{qq}};
\end{displaymath}
the scale dependence of the second term is the same as of the term proportional to $b_{01}$ in the Lagrangian cross correlation only when $\alpha=2$.  But even in this case, there is additional scale dependence from 
\be
\frac{s_1^{pq}}{s_1^{qq}} = 1 + \frac{5+n}{4} \Bigl[1 - (R_p/R_q)^2\Bigr]
 \qquad ({\rm when}\ \alpha=2).
\ee
Whereas filters with $\alpha\ne 2$ are possible, they are slightly artificial, since we typically expect $W$ to be a Taylor series in $(kR)^2$.  Therefore, the $k$-dependence of $2\, {\rm d}W/{\rm d}s_0^{pp}$ will generically differ from $k^2/s_1^{pp}$, so we expect $b_{01}\to b_{01}-1$ to only hold approximately.  Since all filters will have  ${\rm d}W/{\rm d}s_0^{pp}\propto k^2$ at lowest order, 
 $b_{01}\to b_{01}-1$ is a useful approximation for building intuition, but it is not exact.  

\subsection{Excursion set peaks and troughs}\label{espapprox}
The mass fraction in Excursion Set Peaks of height $\nu = \delta_c/\sigma_0^{pp}$ is 
\be
 f_{\rm ESP}(\nu) =  \frac{m/\bar\rho}{(2\pi R_\ast^2)^{3/2}}\,
      \frac{{\rm e}^{-(\nu + |\delta_c|/q_c)^2/2}}{\sqrt{2\pi}}
      \frac{G_1(|\delta_c|/q_c,\gamma_p,\gamma_p\nu)}{\gamma_p\nu}\,,
\label{vfv-esp}
\ee
where 
\be
 R_\ast \equiv \sqrt{3\,s_1^{pp}/s_2^{pp}} \qquad {\rm and}\qquad  
 \gamma_p \equiv s_1^{pp}/\sqrt{s_0^{pp} s_2^{pp}}\,,
\label{VstRstgam}
\ee
and $G_1$ is defined by
\be
 G_n(\beta,\gamma_p,y) \equiv \int_0^\infty {\rm d}x\,x^n\,F(x+\beta\gamma_p)\,
  \frac{\exp[-(x-y)^2/2(1-\gamma_p^2)]}{\sqrt{2\pi(1-\gamma_p^2)}}
 \label{Gn}
\ee
for some $F(x)$ and $-1\le \gamma_p\le 1$ which we will specify shortly.  Note that 
\be
 \frac{\dd G_n}{\dd y} = \frac{G_{n+1}- yG_n}{1-\gamma_p^2}.
\ee
The associated large scale linear bias factor satisfies
\be
 \delta_c\, b_{10}\equiv -\frac{\partial \ln\nu f_{\rm ESP}(\nu)}{\partial\ln\nu} 
   = \nu^2 + \nu\beta
           - \Gamma^2\nu^2 \left(\frac{G_2/G_1}{\gamma\nu} - 1\right)
\label{b1-esp}
\ee
At large $\nu\gg 1$ $G_1\to \gamma\nu\,G_0$ and $G_2/G_0 - (G_1/G_0)^2$ is what $\beta^{-1}$ in approximation (6.20) of BBKS represents.  Their (6.18) shows that their $\beta\to (1-\gamma^2)^{-1}$ when $\gamma\nu\gg 1$ so that 
\begin{equation}
 b_{\rm L} \to \frac{\nu^2 + \nu\beta}{\delta_c}\qquad {\rm when}\quad \nu\gg 1.
 \label{bpeak}
\end{equation}

The upcrossing approximation in the excursion set approach has $F(x)=1$.  In this case, the integrals for the $G_n$ can be done analytically.  The bias factor is 
\be
 \delta_c b_{10} = \nu^2-1 + \frac{{\rm e}^{-\Gamma^2\nu^2/2}/\sqrt{2\pi}\Gamma\nu}{{\rm erfc}(-\Gamma\nu/\sqrt{2})/2 + {\rm e}^{-\Gamma^2\nu^2/2}/\sqrt{2\pi}\Gamma\nu},
 \label{b10up}
\ee
where 
\be
 \Gamma^2 \equiv \frac{\gamma_p^2}{1-\gamma_p^2},
 \ \ \gamma_p^2 \equiv \frac{1}{4s_0^{pp}\,\avg{\!\delta_p'\delta_p'\!}}
 \ \ {\rm and}\ \ \delta_p'\equiv \frac{\dd\delta_p}{\dd s_0^{pp}}
 \label{gammaup}
\ee 
\citep{mps12}.  We use this to produce Figure~\ref{fig:dvoidTH}.

However, for peaks, 
\begin{align}
 F(x) &= \frac{x^3-3x}{2}
  \left\{{\rm erf}\left(x\sqrt{\frac{5}{2}}\right) + {\rm erf}\left(x\sqrt{\frac58}\right)\right\} \nonumber\\
 &\phantom{x^3-3x}
  + \sqrt{\frac{2}{5\pi}}\bigg[\left(\frac{31x^2}{4}+\frac{8}{5}\right){\rm e}^{-5x^2/8} \nonumber\\
&\phantom{\sqrt{x^3-3x+\frac2{5\pi}}[]} + \left(\frac{x^2}{2}-\frac{8}{5}\right){\rm e}^{-5x^2/2}\bigg]\,,
\label{Fxbbks}
\end{align}
(equations~A14--A19 in BBKS).  The presence of the error functions in the definition of $F$ means that the integrals cannot be done analytically in a convenient closed form.



\label{lastpage}

\end{document}